\documentclass[fleqn,usenatbib]{mnras}

\usepackage{newtxtext,newtxmath}

\usepackage[T1]{fontenc}

\DeclareRobustCommand{\VAN}[3]{#2}
\let\VANthebibliography\thebibliography
\def\thebibliography{\DeclareRobustCommand{\VAN}[3]{##3}\VANthebibliography}

\usepackage{graphicx}	
\usepackage{amsmath}	

\usepackage{hyperref}
\hypersetup{
    colorlinks=true,
    linkcolor=cyan,
    citecolor=cyan,
    urlcolor=cyan,
    }

\usepackage{orcidlink}

\usepackage{placeins}

\title[Symbiotic binaries in the \textit{Gaia} data II.]{Symbiotic binaries in the \textit{Gaia} data. II. Symbiotic candidates from general variability classification in DR3
}

\author[J. Merc et al.]{
J. Merc,$^{1,2}$\thanks{E-mail: jaroslav.merc@mff.cuni.cz}\orcidlink{0000-0001-6355-2468}
L. Mulato,$^{3}$\orcidlink{0000-0002-6822-9368}
J.~Miko\l{}ajewska,$^{4}$\orcidlink{0000-0003-3457-0020}
P. G. Beck,$^{2,5}$\orcidlink{0000-0003-4745-2242}
A.~Escorza,$^{2,5}$\orcidlink{0000-0003-3833-2513}
M.~Abdul-Masih,$^{2,5}$\orcidlink{0000-0001-6566-7568}
\newauthor
S. Charbonnel,$^{3}$\orcidlink{0009-0006-0817-4699}
O. Garde,$^{3}$\orcidlink{0000-0002-7850-8360}
P. Le Dû,$^{3}$\orcidlink{0000-0003-2385-0967}
T. Petit$^{3}$
\\
$^{1}$Astronomical Institute, Faculty of Mathematics and Physics, Charles University, V Hole\v{s}ovi\v{c}k{\'a}ch 2, 180 00 Prague, Czech Republic\\
$^{2}$Instituto de Astrof\'isica de Canarias, Calle Vía Láctea, s/n, E-38205 La Laguna, Tenerife, Spain\\
$^{3}$Southern Spectroscopic Project Observatory Team (2SPOT), 45, Chemin du Lac 38690 Châbons, France\\
$^{4}$Nicolaus Copernicus Astronomical Center, Polish Academy of Sciences, Bartycka 18, 00–716 Warsaw, Poland\\
$^{5}$Departamento de Astrof\'isica, Universidad de La Laguna, E-38206 La Laguna, Tenerife, Spain
}

\date{Accepted 2026 July 14. Received 2026 July 06; in original form 2026 June 01}

\pubyear{\the\year{}}

\begin{document}
\label{firstpage}
\pagerange{\pageref{firstpage}--\pageref{lastpage}}
\maketitle

\begin{abstract}
We investigate the reliability of the symbiotic star class in the general variability classification of \textit{Gaia} DR3 and search for new genuine symbiotic systems among the objects assigned to this category. The sample contains 649 sources, including 246 previously known symbiotic stars, 61 literature candidates, and 339 new candidates proposed by the \textit{Gaia} variability pipeline. Diagnostics based on the \textit{Gaia} colour-magnitude diagram, near-infrared photometry, and the pseudo-equivalent width of H$\alpha$ indicate that a large fraction of the new candidates are likely contaminants, predominantly pulsating red giants. To quantify the contamination, we constructed a Random Forest classifier trained on confirmed symbiotic stars and on Mira and semi-regular variables, using \textit{Gaia} photometry, variability parameters, H$\alpha$ measurements from XP spectra, and infrared colours. The classifier reaches a balanced accuracy of $\approx$0.94 and efficiently separates most symbiotic binaries from single evolved stars, leaving only eight strong candidates among the 339 newly proposed objects. Follow-up spectroscopy confirms three new symbiotic stars through the presence of high-excitation emission lines, while several additional objects remain possible symbiotics. Our results show that the \textit{Gaia} DR3 variability classification efficiently recovers known symbiotic stars but has low purity due to overlap with pulsating red giants. The small number of newly confirmed systems implies that the discrepancy between predicted and observed Galactic symbiotic populations remains unresolved, although \textit{Gaia} provides a powerful basis for future searches combining variability, spectroscopic indicators, and multi-wavelength data.
\end{abstract}




\section{Introduction}
Symbiotic stars are a class of long-period interacting binaries in which an evolved giant transfers material to a hot, compact companion, most commonly a white dwarf. The donor is typically a red giant branch (RGB) or asymptotic giant branch (AGB) star, while the accretor ionizes the surrounding circumstellar material, giving rise to the rich and complex observational phenomenology that characterizes these systems (see reviews by \citealt{2012BaltA..21....5M}; \citealt{2019arXiv190901389M}; \citealt{2025Galax..13...49M}). One of the long-standing open problems in the field is the large discrepancy between the predicted and observed number of symbiotic stars in the Milky Way. Population estimates range from a few thousand to several hundred thousand systems \citep[e.g., ][]{2003ASPC..303..539M,2006MNRAS.372.1389L, 2025A&A...698A.155L}, whereas only $\sim$300 symbiotic stars are currently known \citep[][]{2019RNAAS...3...28M,2019AN....340..598M,2026arXiv260625623M,2019ApJS..240...21A}. This large discrepancy motivates systematic searches for new symbiotic stars using modern large-scale ground- and space-based surveys.

The European Space Agency’s \textit{Gaia} mission \citep{2016A&A...595A...1G} has revolutionised many areas of astrophysics. In this series of papers, we investigate the impact of the third data release (\textit{Gaia} DR3; \citealt{2023A&A...674A...1G}) on the study of symbiotic stars. In the first paper of the series \citep[][]{2026MNRAS.546ag125M}, we focused on confirmed symbiotic systems and examined how they are represented in the \textit{Gaia} data, as well as which of the parameters published in DR3 can be reliably used in studies of the symbiotic population.

In this follow-up paper, we focus on the results of the \textit{Gaia} DR3 variability classification, which, among other classes, included symbiotic stars. In Section \ref{sec:variability}, we briefly describe the supervised machine-learning procedure used to select candidates, examine the training set adopted in the classification pipeline, and discuss the 649 objects classified as symbiotic by this algorithm. We also perform a first comparison of this "\textit{Gaia} DR3 symbiotic sample" with the population of known symbiotic stars. In Section \ref{sec:machine_learning}, we develop our own machine-learning classifier based on the Random Forest method, which combines \textit{Gaia} photometry with spectroscopic indicators and data from additional surveys in order to identify contaminants in the \textit{Gaia} symbiotic sample. In Section \ref{sec:our_candidates}, we analyse the objects that were not known as symbiotic stars prior to \textit{Gaia} DR3 and evaluate which of them are likely genuine new members of the class. In Section \ref{sec:akras}, we compare our results with those of \citet{2026MNRAS.546ag105A}{, who analysed the same \textit{Gaia} DR3 sample using empirical photometric selection criteria based on infrared colours and H$\alpha$ information,} and we discuss some caveats of their work. The main conclusions are summarised in Section \ref{sec:conclusions}.

\section{General variability classification in DR3}\label{sec:variability}
The general variability classification in \textit{Gaia} DR3 was performed using supervised machine-learning algorithms \citep[][]{2023A&A...674A..14R,2023A&A...674A..13E}. The classifiers were trained to assign variable sources to 25 variability classes. Multiple classifier configurations were employed (in total, more than 100 classifiers, trained with two machine learning algorithms{, namely Distributed Random Forest and eXtremeGradient Boosting,} contributed to the final classification), with several classifiers evaluated for each variability class. The outputs of the individual classifiers were subsequently combined, and only the final aggregated solution is reported in the \texttt{vari\_classifier\_result} table.

The classifiers were trained using photometric, astrometric, and time-series attributes derived from the $G$, $BP$, and $RP$ FoV-transit magnitudes. These include measures of variability (e.g., standard deviation, skewness, Stetson index), correlations between bands, colours, parallax-based luminosities, and periodogram features (see Appendix B in \citealt{2023A&A...674A..14R}).

The success of this approach depends on how well the variability classes can be separated using this set of features (e.g., no spectroscopic parameters were used), and, being a supervised method, also on the quality and representativeness of the training set. Classes that are poorly represented in the training data or intrinsically overlap in the feature space are more difficult to identify reliably, which can lead to misclassifications. In the following subsection, we examine the training set used for symbiotic stars in \textit{Gaia} DR3, and in subsequent sections, we discuss the properties of the sources classified as SYST, including both previously known symbiotic stars and new candidate objects.

\subsection{Training set}
The training sample of symbiotic stars used in the \textit{Gaia} DR3 variability classification originates from a cross-match of literature compilations, primarily the catalogues of \citet{2000A&AS..146..407B} and \citet{2019ApJS..240...21A}. The resulting cross-match table, published by \citet{2023A&A...674A..22G}, contains 361 objects identified as symbiotic stars. The training set used for the variability classifier was subsequently selected from this cross-matched sample.

A closer inspection of this complete sample reveals several issues with the adopted \textit{Gaia} counterparts. Six objects have incorrect \textit{Gaia} identifications, while 27 symbiotic stars appear twice in the catalogue. In many cases, this duplication arises from imprecise literature coordinates, most commonly from \citet{2000A&AS..146..407B}. When the same object was later listed by \citet{2019ApJS..240...21A} with improved coordinates, the crossmatch algorithm could associate the two entries with different \textit{Gaia} sources. As a consequence, the same symbiotic star appears twice in the crossmatch table with distinct \textit{Gaia} source identifiers, and the duplication was not removed because the entries correspond to different \textit{Gaia} objects. In some cases, the matching procedure also appears to have favoured a faint anonymous source over the correct bright counterpart. This likely reflects the fact that the crossmatch considered not only positional agreement but also literature magnitudes; if the adopted literature magnitude was inaccurate or underestimated, the algorithm could select a fainter source located slightly further from the reported position.

In addition, the sample mixes confirmed symbiotic stars and candidate systems. According to the classification adopted in the New Online Database of Symbiotic Variables \citep[NODSV;][]{2019RNAAS...3...28M,2019AN....340..598M,2026arXiv260625623M}, only 260 of the 361 objects ($\sim$72\%) can be considered confirmed symbiotic stars. Another 64 objects ($\sim$18\%) have only candidate status, while 37 objects ($\sim$10\%) are not symbiotic stars at all. The latter category is largely explained by incorrect \textit{Gaia} counterpart identifications, which account for 33 of these 37 cases.

The subset of objects used to train the \textit{Gaia} DR3 variability classifier is somewhat cleaner. According to \citet{2023A&A...674A..14R}, 316 objects from the original list were used in the training set. Because the exact list of objects is not published, we attempted to approximate the training selection using mainly the criteria described in Section 3.1.2 of \citet{2023A&A...674A..14R}. Specifically, we excluded objects with fewer than five G-band observations, sources lacking BP and RP magnitudes, and objects not classified as variables in \textit{Gaia} DR3. Applying these criteria yields a sample of 318 objects, which almost matches the number reported for the training set.

These filters remove most of the incorrectly included objects, but the resulting sample still contains a mixture of confirmed symbiotic stars and candidates. In the classification from NODSV, 252 of the 318 objects ($\sim$79\%) are confirmed symbiotic systems, 60 objects ($\sim$19\%) are candidates, and 6 objects ($\sim$2\%) are non-symbiotic stars. The presence of likely non-symbiotic sources is also evident, for example, in the \textit{Gaia} CMD \citep[see Fig. E.64 in][]{2023A&A...674A..14R}, where some objects occupy the main-sequence or subgiant region rather than the giant branch where symbiotic systems are expected \citep[][]{2026MNRAS.546ag125M}. The composition of the training sample, therefore, remains imperfect, containing not only confirmed symbiotic stars but also a substantial fraction of candidates and a small number of non-symbiotic objects. Such heterogeneity may propagate into the resulting classifier and affect the purity of the identified symbiotic candidates in \textit{Gaia} DR3. 

\subsection{Output sample of symbiotic stars}\label{sec:sample}
In total, 649 objects are classified as symbiotic stars in \textit{Gaia} DR3, that is, sources for which {\tt best\_class\_name} is "SYST" in the {\tt vari\_classifier\_result} table. It should be noted that small classes, including symbiotic stars, were protected against reassignment to alternative variability classes when a source could plausibly belong to multiple categories. As a result, some objects may have a higher probability of belonging to another variability type, yet are still classified as symbiotic stars in the final catalogue. As already noted by \citet{2023A&A...674A..14R}, the most common false positives in the symbiotic sample are objects classified in the training sets as long-period variables (LPVs), semi-regular pulsators, OGLE small-amplitude red giants (OSARGs), or Miras (see their Table~3). 

A large fraction of the training sources are classified as SYST in the final catalogue (219 objects), while 49 are reported as false negatives in \citet{2023A&A...674A..14R}. The status of the remaining training-set sources is not explicitly discussed. Consequently, the true positive rate (and the performance of the classifier) is difficult to interpret, since many of the recovered objects were part of the training sample, and it is not clear whether cross-validation was employed during the training procedure.

The published {\tt best\_class\_score} does not represent a calibrated probability but a normalised rank derived from the classifier outputs. Following \citet{2023A&A...674A..14R}, posterior probabilities from individual classifiers were sorted within each class and converted to normalised ranks. The final score, therefore, reflects the relative position of a source among the objects of the class rather than the absolute probability of belonging to a given variability type.

We cross-matched the \textit{Gaia} DR3 symbiotic sample with NODSV. Of the 649 objects, 246 correspond to previously known symbiotic stars (see the note on AS 269 in Appendix \ref{app:as269}), while 61 were already listed as candidates prior to \textit{Gaia} DR3. Three sources are listed in NODSV as misclassified objects: 4U~1954+31 \citep[][]{2020ApJ...904..143H}, now classified as a high-mass X-ray binary, Hen~3-1383, a luminous blue variable \citep[][]{2012MNRAS.421.3325G}, and Gaia DR3 406134544052580377, previously incorrectly identified as a counterpart of a known symbiotic star Terz V 2513 \citep[][]{2026MNRAS.545f2094M}. All three objects were included in the training sample because they were previously listed as known symbiotic stars or candidates in \citet{2019ApJS..240...21A}. The remaining 339 objects represent new symbiotic candidates that had not been previously proposed as such.

Our primary goal is to investigate the nature of the newly proposed candidates. Nevertheless, we analyse the properties of the entire \textit{Gaia} DR3 symbiotic sample in order to place these candidates in the broader context of the population identified by the classifier. For this purpose, we typically divide the sample of 649 objects into four groups: confirmed symbiotic stars listed in NODSV, candidate symbiotic stars listed in NODSV, new candidates that are not included in NODSV, and objects previously considered as symbiotic stars, but reclassified in the literature (i.e., misclassified sources; not shown in most of the plots for simplicity).

The confirmed symbiotic stars span almost the entire range of {\tt best\_class\_score} (with about 55\% having {\tt best\_class\_score} $\geq 0.5$), indicating that they are not generally ranked among the strongest candidates by the classifier. Naively, this suggests that the sample of new candidates may contain a substantial number of promising sources that could merit further follow-up observations.

\begin{figure}
\centering
\includegraphics[width=\columnwidth]{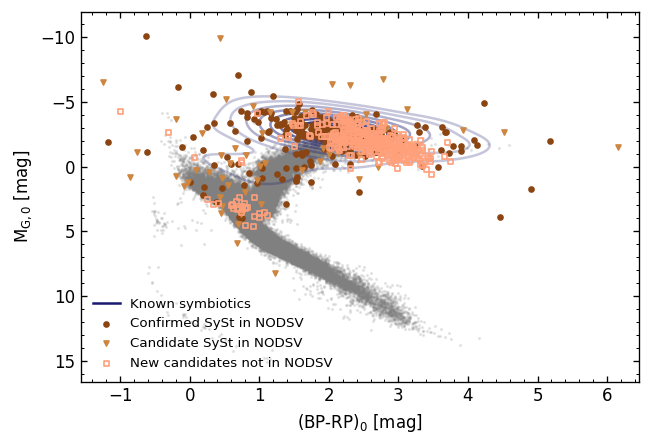}
\caption{Extinction-corrected \textit{Gaia} colour-magnitude diagram of the objects analysed in this work. Absolute magnitudes $M_G$ and dereddened \mbox{$(BP-RP)_0$} colours were computed using geometric distances from \citet{2021AJ....161..147B} and extinction estimates derived with the {\tt mwdust} code from combined 3D dust maps (see the text for the details). Contours show the distribution of confirmed symbiotic stars from NODSV, {irrespective of their \textit{Gaia} DR3 classification. The symbols denote the subset of \textit{Gaia} DR3 sources classified as SYST, separated into confirmed symbiotic stars in NODSV, candidate symbiotic stars in NODSV, and new candidates.}. The background shows a well-characterized sample of \textit{Kepler} stars from \citet{2025A&A...696A.243G}.}
\label{fig:gaia_cmd}
\end{figure}

In Fig.~\ref{fig:gaia_cmd}, we show the positions of all objects in the \textit{Gaia} colour-magnitude diagram (CMD). To calculate extinction-corrected $BP-RP$ colors and absolute magnitudes $M_G$, we adopted geometric distances from \citet{2021AJ....161..147B} and extinction values derived with the {\tt mwdust} code \citep{2016ApJ...818..130B}, based on combined 3D dust maps from \citet{2003A&A...409..205D}, \citet{2006A&A...453..635M}, and \citet{2019ApJ...887...93G}, following the same procedure as in our analysis of known symbiotic stars \citep{2026MNRAS.546ag125M}. The distribution of the known symbiotic sample (all confirmed systems from NODSV, not only those classified as SYST in \textit{Gaia} DR3) is shown as contours.

It is evident that a large fraction of the candidates occupy the same region of the CMD as the confirmed symbiotic stars, although some lie closer to the main sequence or subgiant branch. A few known symbiotic stars appear as outliers in this diagram, mostly dusty systems affected by additional extinction that is not fully corrected \citep[see more details in][]{2026MNRAS.546ag125M}. Overall, given that the CMD position was in principle one of the features used in the general variability classification pipeline of \textit{Gaia} DR3, this result is not unexpected. However, aside from potentially excluding a few poorly matching candidates, the CMD position alone provides limited constraints on the nature of the majority of candidates.

\begin{figure}
\centering
\includegraphics[width=\columnwidth]{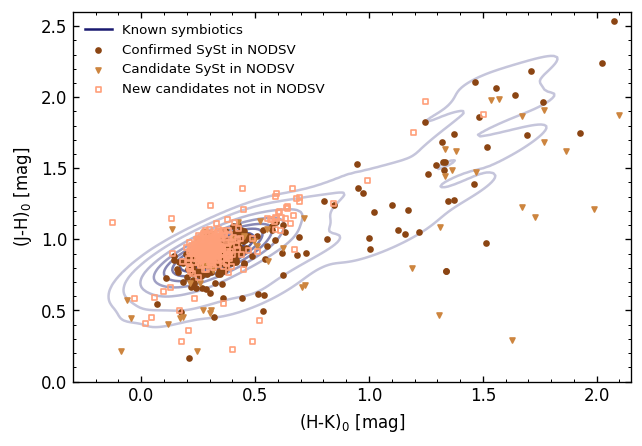}
\caption{2MASS near-infrared colour--colour diagram of the objects analysed in this work. Contours show the distribution of all confirmed symbiotic stars from NODSV for comparison{, as in Fig. \ref{fig:gaia_cmd}.}}
\label{fig:ir_color_color}
\end{figure}

In Fig.~\ref{fig:ir_color_color}, we show the positions of the candidates in the 2MASS near-IR colour-colour diagram \citep[][]{2006AJ....131.1163S}, a diagnostic commonly used to distinguish symbiotic stars from other emission-line objects \citep[see, e.g.,][]{2001A&A...377L..18S,2007MNRAS.376.1120P,2008A&A...480..409C,2019MNRAS.483.5077A,2021MNRAS.506.4151M,2025MNRAS.542.3097B}. This diagram is completely independent of the \textit{Gaia} DR3 classification, as near-IR data were not used in the classifier. It is apparent that the candidates occupy the region of the 2MASS diagram populated by the bulk of the known symbiotic population, with the upper-right tail corresponding to dusty symbiotic systems. Based on this diagram, the sample does not appear to be strongly contaminated by typical emission-line interlopers such as planetary nebulae, Be stars, or young stellar objects. This is not unexpected, however, because the classifier primarily selected objects containing red giants, which occupy a similar region in the near-IR colour-colour space.

\begin{figure}
\centering
\includegraphics[width=\columnwidth]{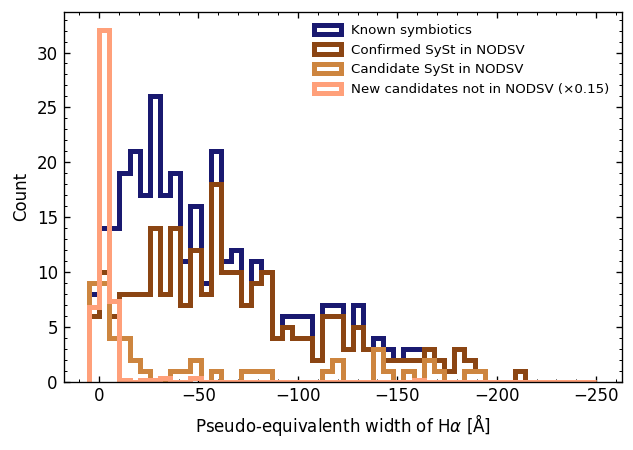}
\caption{Distribution of the pseudo-equivalent width of H$\alpha$ (pEW H$\alpha$) for objects classified as SYST in \textit{Gaia} DR3. The distribution of confirmed symbiotic stars from NODSV is shown for comparison{, as in Fig. \ref{fig:gaia_cmd}.}}
\label{fig:halpha}
\end{figure}

As mentioned earlier, the most common false positives in the \textit{Gaia} DR3 symbiotic sample, identified through comparison with the training set, are pulsating red giants. This is expected, and was previously noted by \citet{2025Galax..13...49M}, because the variability of symbiotic systems, either quasi-sinusoidal due to orbital motion or arising from the red giant component, can mimic pulsation signatures of single pulsating giants, and the associated periodicities and positions in the CMD are similar (see also Sect.~4.2 and Fig.~4 of \citealt{2026MNRAS.546ag125M}). \citet{2026MNRAS.546ag125M} showed, however, that symbiotic stars can be distinguished by their emission-line nature, and the pseudo-equivalent width of H$\alpha$ (pEW H$\alpha$) published in \textit{Gaia} DR3 is useful in this regard. In Fig.~\ref{fig:halpha}, we show the distribution of pEW H$\alpha$ for the \textit{Gaia} DR3 candidates and compare it with that of the known symbiotic systems. When restricting to the 339 previously unidentified candidates (i.e., omitting known symbiotic stars and previously proposed candidates; lightest line in the histogram), the distribution is markedly different: the majority of these new candidates do not show significant H$\alpha$ emission and thus differ from most known symbiotic stars. Based on this comparison, we conclude that a large fraction of the 339 candidates are likely contaminants, primarily non-symbiotic pulsating red giants, a point we explore further in the following sections.

\section{Machine-learning analysis of the sample}\label{sec:machine_learning}

The simple comparisons presented in the previous sections suggest that a significant fraction of the newly identified \textit{Gaia} DR3 candidates are likely contaminants, primarily pulsating red giants. This conclusion is mainly supported by the absence of detectable H$\alpha$ emission in the \textit{Gaia} XP spectra for many objects. Such behaviour is inconsistent with the majority of shell-burning symbiotic stars, which host luminous, hot white dwarfs undergoing quasi-steady hydrogen burning. These systems show rich emission-line spectra \citep[see, e.g.][]{2019arXiv190901389M,2025Galax..13...49M}, and, as demonstrated by \citet{2026MNRAS.546ag125M}, their H$\alpha$ emission is readily detectable even in the low-resolution \textit{Gaia} XP spectra.

On the other hand, not all symbiotic stars are expected to show strong emission lines. In accretion-powered systems, whose luminosity is dominated by accretion rather than nuclear burning, emission features can be weak or even absent. Such objects are often identified through ultraviolet excess, flickering variability associated with accretion discs, or X-ray emission \citep[e.g.,][]{2013A&A...559A...6L,2016MNRAS.461L...1M,2019arXiv190901389M}. However, none of these diagnostics was used as input features in the \textit{Gaia} DR3 variability classification pipeline. Therefore, there is no strong reason to expect that the DR3 symbiotic sample should contain a large number of accretion-powered systems. Given the input data available to the classifier, objects without detectable emission lines are more likely to be pulsating red giants than genuine symbiotic binaries.

Motivated by this, we perform a more systematic analysis of the \textit{Gaia} DR3 sample using a supervised machine-learning classifier. The goal is to evaluate whether the newly proposed candidates resemble known symbiotic stars or typical contaminants in a multidimensional parameter space defined by photometric, spectroscopic, and variability-related features. In particular, we focus on separating symbiotic systems from pulsating red giants, which represent the dominant expected source of contamination in the \textit{Gaia} DR3 symbiotic sample. In the following subsections, we describe the construction of the training sets and the methodology used to build and evaluate the classifier.

\subsection{Training sets and comparison of populations}

In order to evaluate the nature of the \textit{Gaia} DR3 symbiotic candidates, we constructed training samples representing both confirmed symbiotic stars and the most likely contaminants. As a reference sample of genuine symbiotic systems, we used objects listed (as confirmed) in the NODSV \citep[][]{2019RNAAS...3...28M,2019AN....340..598M,2026arXiv260625623M}. These objects were cross-matched with the \textit{Gaia} DR3, 2MASS, and AllWISE \citep[][]{2010AJ....140.1868W} catalogues, and only those with counterparts in all three surveys were retained. We further required that the sources have pEW measurements of H$\alpha$ from the XP spectra, are classified as variables in \textit{Gaia} DR3, and have all variability-related parameters used in this work available. In addition, we excluded sources without geometric distances from \citet{2021AJ....161..147B}, which we used for the computation of extinction-corrected magnitudes and colours in a homogeneous way. 

The dominant expected source of contamination in the \textit{Gaia} DR3 symbiotic sample are pulsating red giants, in particular Mira and semi-regular (SR) variables, whose variability properties and positions in the colour--magnitude diagram are similar to those of symbiotic systems. To construct a representative comparison sample of such objects, we used the International Variable Star Index (VSX; \citealt{2006SASS...25...47W}). We randomly selected 10\,000 objects classified as Mira and 10\,000 SR variables and cross-matched them with the \textit{Gaia} DR3, 2MASS, and AllWISE catalogues. 

The same selection criteria as for the symbiotic-star sample were then applied. We retained only objects with counterparts in all required surveys, with available H$\alpha$ pEW measurements, and with all variability-related parameters present in \textit{Gaia} DR3. In addition, we also required that the sources are classified as long-period variables in the \textit{Gaia} DR3 LPV SOS pipeline \citep[][]{2023A&A...674A..15L}, in order to minimise the risk of including misclassified objects. Finally, only sources with geometric distances from \citet{2021AJ....161..147B} were kept, so that extinction-corrected photometric parameters can be derived consistently.

\begin{figure*}
\centering
\includegraphics[width=\textwidth]{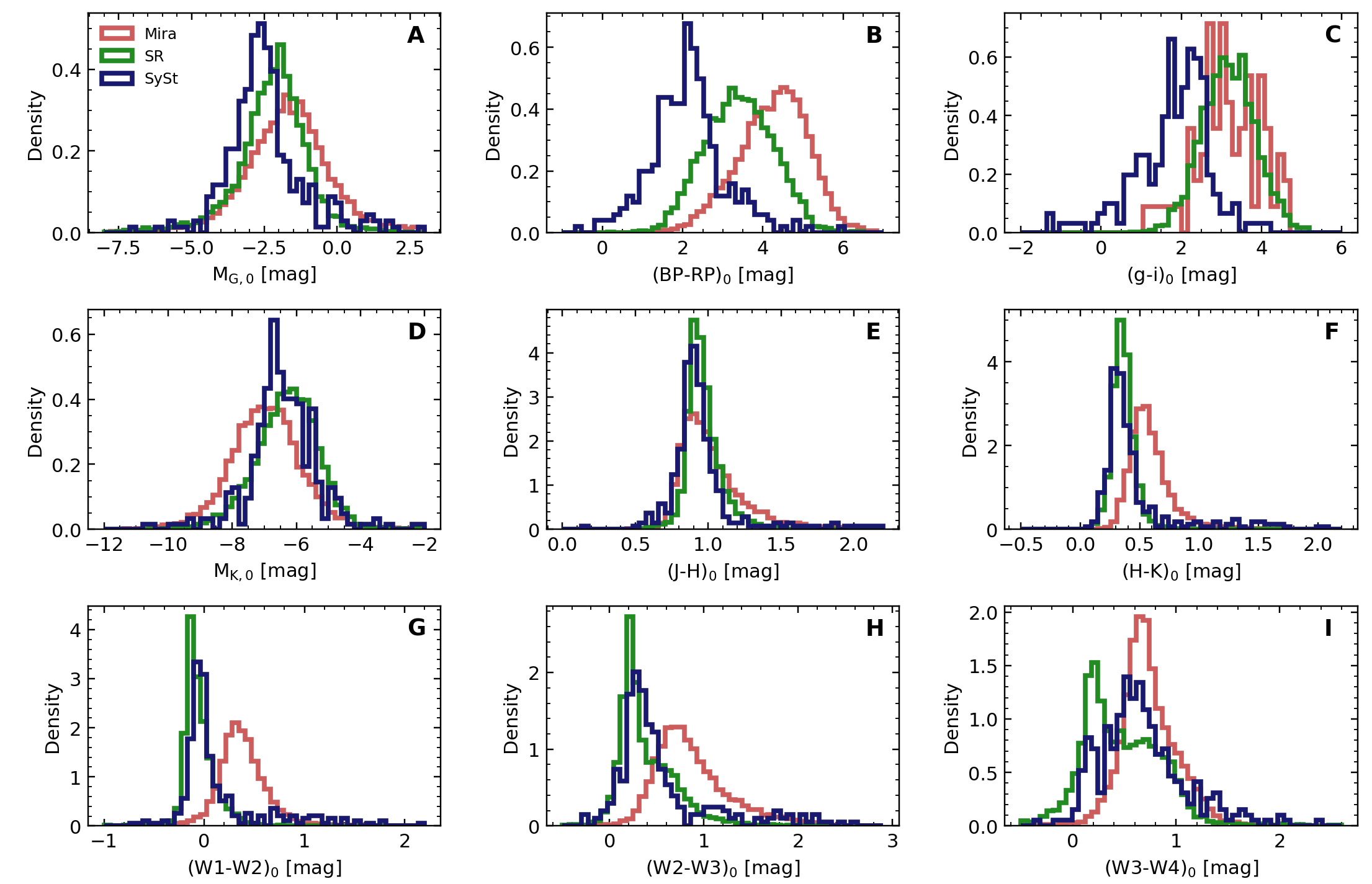}

\caption{Comparison of parameter distributions for confirmed symbiotic stars (blue), SR variables (green), and Mira variables (red). Histograms show normalised densities.
\textbf{A:} Extinction-corrected absolute $G$ magnitude.
\textbf{B:} Dereddened $(BP-RP)_0$ color.
\textbf{C:} Synthetic \textit{Gaia} DR3 $g-i$ color.
\textbf{D:} Absolute $K$-band magnitude.
\textbf{E:} Dereddened $(J-H)_0$ color.
\textbf{F:} Dereddened $(H-K)_0$ color.
\textbf{G:} Dereddened WISE $(W1-W2)_0$ color.
\textbf{H:} Dereddened WISE $(W2-W3)_0$ color.
\textbf{I:} Dereddened WISE $(W3-W4)_0$ color.}
\label{fig:features_1}
\end{figure*}

\begin{figure*}
\centering
\includegraphics[width=\textwidth]{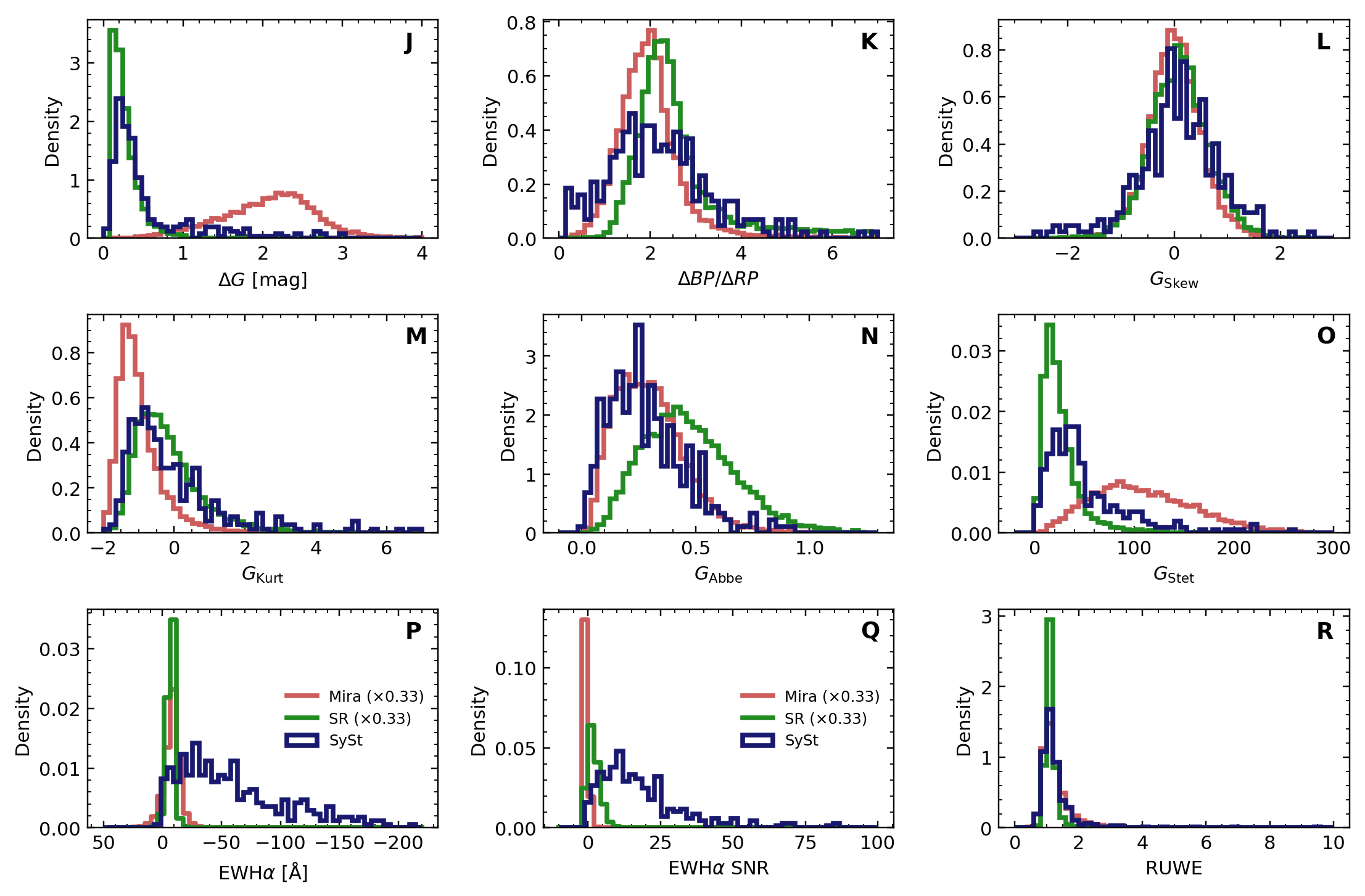}

\contcaption{Variability and spectroscopic parameters.
\textbf{J:} Trimmed $G$-band variability amplitude ($\Delta G$).
\textbf{K:} Ratio of variability amplitudes in $BP$ and $RP$ bands.
\textbf{L:} Skewness of the $G$-band light curve.
\textbf{M:} Kurtosis of the $G$-band light curve.
\textbf{N:} Abbe value of the $G$-band light curve.
\textbf{O:} Stetson variability index.
\textbf{P:} Pseudo-equivalent width of H$\alpha$.
\textbf{Q:} Signal-to-noise ratio of the H$\alpha$ measurement.
\textbf{R:} Renormalised unit weight error (RUWE). 
For clarity, the Mira and SR distributions in panels P and Q are scaled by a factor of 0.33. All histograms show normalised densities.}
\label{fig:features_2}
\end{figure*}

In Fig.~\ref{fig:features_1}, we compare the distributions of selected parameters for confirmed symbiotic stars, Mira variables, and SR variables using histograms. Fig.~\ref{fig:features_1}A shows the distribution of extinction-corrected absolute $G$ magnitudes. Symbiotic stars are, on average, the brightest of the three groups, most likely because the nebular emission contributes significantly to the flux in the $G$ band in addition to the red giant itself. However, the distributions strongly overlap, indicating that this parameter alone cannot be used for a reliable separation of the populations.

The effect of the nebular contribution becomes more evident when dereddened optical colours are compared. Fig.~\ref{fig:features_1}B shows the distribution of $(BP-RP)_0$, while Fig.~\ref{fig:features_1}C presents the comparison of synthetic \textit{Gaia} DR3 $g$ and $i$ magnitudes \citep[][]{2023A&A...674A..33G}. Although symbiotic stars contain the same types of pulsating giants as the comparison sample (with Mira variables being less represented among symbiotic systems), the presence of nebular emission makes symbiotic stars systematically bluer. In the $(BP-RP)_0$ distribution, SR and Mira variables are also partially separated, reflecting their different evolutionary stages, with Mira variables typically being more evolved, systematically cooler, and therefore redder.

Fig.~\ref{fig:features_1}D shows the comparison of absolute $K$-band magnitudes. In this case, the distributions of symbiotic stars, SRs, and Miras largely overlap, although the most luminous objects are predominantly Mira variables, consistent with their more advanced evolutionary state and, in many cases, the presence of circumstellar dust. In the near-infrared, SR variables are difficult to distinguish from symbiotic stars, because the additional components present in symbiotic systems contribute only weakly at these wavelengths.

Fig.~\ref{fig:features_1}E--I show the comparison of near- and mid-infrared colours. The $J-H$ distribution (Fig.~\ref{fig:features_1}E) is very similar for all three groups, while differences become more visible in $H-K$ (Fig.~\ref{fig:features_1}F) and in the WISE colours (Fig.~\ref{fig:features_1}G--I), where Mira variables tend to be redder due to the presence of warm circumstellar dust.

Overall, these comparisons demonstrate that the bulk of the symbiotic-star population can be statistically separated from single pulsating red giants when optical and infrared magnitudes and colours are considered simultaneously, although no single parameter provides a clean discrimination between the classes.

Fig.~\ref{fig:features_1}J--O focus on the variability-related parameters provided in \textit{Gaia} DR3. Fig.~\ref{fig:features_1}J compares the trimmed $G$-band variability amplitude, defined as the difference between the 95th and 5th percentile of the $G$-band light curve after removing outliers, which provides a robust estimate of the variability range. As expected, Mira variables show the largest amplitudes, while SR variables and symbiotic stars are largely indistinguishable. This is not surprising, because the variability detected by \textit{Gaia} in many symbiotic systems is caused by the same physical mechanism as in SR variables, namely pulsations of the red giant component \citep{2026MNRAS.546ag125M}. 

Fig.~\ref{fig:features_1}K compares the amplitudes of variability in the $BP$ and $RP$ bands. On average, the ratio of the amplitudes differs slightly between the groups, with Mira variables tending to show smaller ratios than SR variables, but the distributions overlap significantly and do not provide a clear separation.

In Fig.~\ref{fig:features_1}L, we compare the skewness of the $G$-band light curves, which describes the asymmetry of the magnitude distribution and can indicate whether the variability is dominated by sharp maxima or minima. The distributions of skewness are very similar for all three groups, indicating that this parameter alone has little discriminating power. Fig.~\ref{fig:features_1}M shows the kurtosis of the $G$-band light curves, which measures the degree of peakedness of the magnitude distribution. Symbiotic stars and SR variables again show very similar distributions, while Mira variables tend to have slightly more negative kurtosis values on average, reflecting their more regular, large-amplitude pulsations.

Fig.~\ref{fig:features_1}N presents the Abbe value, a statistic describing the smoothness of the light curve by comparing point-to-point variations with the overall variance. Lower values correspond to smoother, more correlated variability. Mira variables and symbiotic stars show similar Abbe values, both typically lower than those of SR variables, although the overlap between the distributions remains substantial. Fig.~\ref{fig:features_1}O shows the Stetson variability index, which measures the degree of correlated variability between different photometric bands. Similar to the trimmed $G$-band amplitude, this parameter clearly separates Mira variables, which show the largest variability, while SR variables and symbiotic stars largely overlap.

In Fig.~\ref{fig:features_1}P, we compare the pseudo-equivalent width of H$\alpha$ derived from the \textit{Gaia} XP spectra. The distributions of symbiotic stars and pulsating variables are clearly different. Most Mira and SR variables show very small values consistent with the absence of emission lines, often with large uncertainties. The uncertainties are typically larger for Mira variables, because strong pulsations significantly change the continuum level. Fig.~\ref{fig:features_1}Q shows the signal-to-noise ratio of the H$\alpha$ measurement, which itself provides additional separation between the groups, as genuine emission-line objects tend to have higher signal-to-noise values.

Finally, Fig.~\ref{fig:features_1}R shows the distribution of the renormalised unit weight error (RUWE), which is often used as an indicator of unresolved binarity. As already noted by \citet{2026MNRAS.546ag125M}, RUWE is not a reliable binary indicator for symbiotic stars, and the distributions for symbiotic systems, Mira variables, and SR variables are nearly identical in our sample.

\subsection{Random Forest classifier}\label{sec:rf}

Motivated by the systematic differences between symbiotic stars, Mira variables, and SR variables discussed in the previous subsection, we constructed a supervised machine-learning classifier to separate these populations in a multidimensional parameter space. {Our objective was not to identify the optimal classification algorithm in a general sense, but to develop a robust classifier tailored to the specific task of distinguishing symbiotic stars from pulsating giants, which constitute the dominant contaminants in the candidate sample. We therefore adopted a Random Forest (RF) classifier, a well-established method for heterogeneous tabular data that requires relatively little tuning while providing robust performance and interpretable feature-importance estimates. The classifier was trained} on the confirmed samples of symbiotic stars, Miras, and SR variables using the set of photometric, spectroscopic, and variability-related features described above.

The classifier was implemented using the \texttt{RandomForestClassifier} from the \texttt{scikit-learn} library \citep[][]{2011JMLR...12.2825P}, allowing classification into three classes (Mira, SR, and symbiotic star). The input feature vector consisted of extinction-corrected absolute magnitudes and colours from \textit{Gaia}, 2MASS, and WISE, the pEW of H$\alpha$ and its signal-to-noise ratio from \textit{Gaia} XP spectra, and several variability statistics derived from the \textit{Gaia} light curves, including the trimmed $G$-band amplitude, Abbe value, Stetson index, and the ratio of variability amplitudes in $BP$ and $RP$ bands.

Hyperparameters of the Random Forest were optimised using a grid search with stratified cross-validation. The explored parameter space included the number of trees, maximum tree depth, minimum number of samples per leaf, feature subsampling fraction, and class weighting. The best-performing model used 400 trees, balanced class weights, a maximum depth of 15, and a minimum leaf size of 2. The spread of the scores among the tested hyperparameter combinations was small, indicating that the performance of the classifier is stable with respect to the exact parameter choice.

\subsubsection{Classification performance}

To obtain an unbiased estimate of the classifier performance, we used nested stratified cross-validation. The inner loop was used for hyperparameter optimisation, while the outer loop provided predictions for all objects that were not used during training (80:20 split). This procedure avoids overly optimistic performance estimates and ensures that the reported accuracy reflects the true predictive power of the model. Without cross-validation, the accuracy is close to 100\%; however, this result is not meaningful as an estimate of the true performance, since the classifier remembers the training set. 

The performance was evaluated using precision, recall, F1-score, balanced accuracy, confusion matrices, and receiver operating characteristic (ROC) curves for each class. The overall performance of the classifier is high, with a balanced accuracy of 0.94 obtained from the nested cross-validation. Mira and SR variables are almost perfectly separated, with correct classification rates above 98\% for both classes. Symbiotic stars are recovered with a recall of 0.86, meaning that about 14\% of known symbiotic systems are misclassified.

\begin{figure}
\centering
\includegraphics[width=0.9\columnwidth]{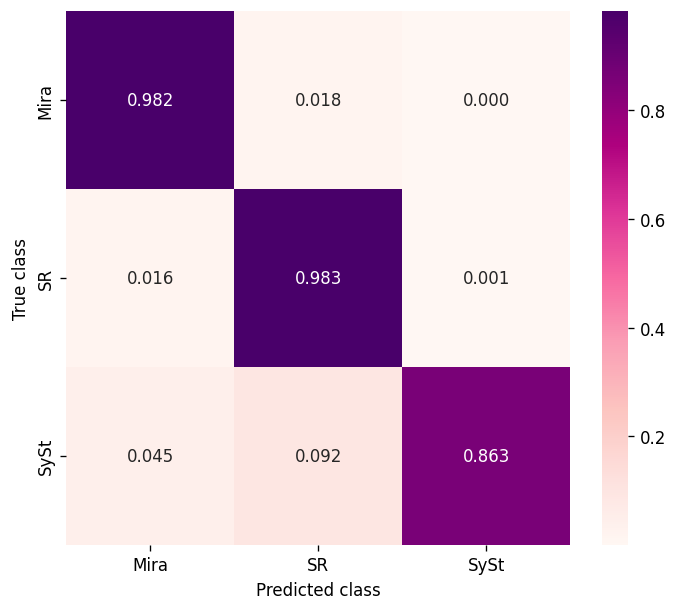}
\caption{Confusion matrix for the RF classifier distinguishing confirmed symbiotic stars from Mira and SR variables. The matrix shows normalised fractions of objects assigned to each class, illustrating the high completeness of the symbiotic star sample and the main source of contamination arising from SR variables, whose properties overlap with those of accreting-only symbiotic systems.}
\label{fig:conf_matrix}
\end{figure}

Inspection of the confusion matrix (Fig. \ref{fig:conf_matrix}) shows that only a negligible fraction of Mira and SR variables are classified as symbiotic stars (see Appendix \ref{app:misclassified}), while most misclassified objects are symbiotic systems assigned to the SR or Mira classes. A total of 43 confirmed symbiotic stars are misclassified in the cross-validation. Examination of these objects shows that they are predominantly accretion-powered systems without strong nebular emission, and therefore their colours and spectra resemble those of normal pulsating red giants (e.g., SU Lyn, \citealt{2016MNRAS.461L...1M}; V420 Hya, \citealt{2013A&A...559A...6L}; V520 And, \citealt{2025AstBu..80...58M}; Gaia23cse, \citealt{2025A&A...702A.269M}). These objects are sometimes not considered among the "typical" members of the symbiotic family. In several cases where the cool component is a Mira variable \citep[e.g, CGCS 6306, ][]{2025A&A...699A.117M}, the variability properties alone are sufficient for the classifier to assign the object to the Mira class. Given the available input features, such behaviour is expected and cannot be avoided without including additional diagnostics such as ultraviolet or X-ray data, which are, however, not available for each source.  

\begin{figure}
\centering
\includegraphics[width=0.9\columnwidth]{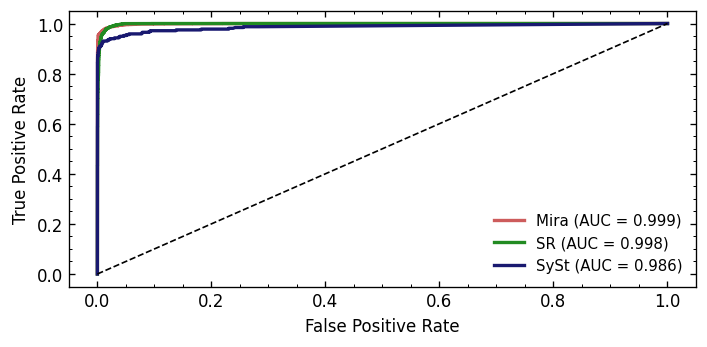}
\caption{ROC curves for the RF classifier. The curves show the true positive rate as a function of the false positive rate. The large area under the curve indicates a good separability between symbiotic stars and single pulsating giants using the selected Gaia variability and photometric parameters.}
\label{fig:roc}
\end{figure}

The ROC curves further confirm the high performance of the classifier, with areas under the curve (AUCs), where values of 1 indicate perfect separability, and 0.5 corresponds to random classification, of 0.999 for Miras, 0.998 for SR variables, and 0.986 for symbiotic stars (Fig.~\ref{fig:roc}). These values indicate excellent separability of the classes in the adopted parameter space.

\subsubsection{Feature importance}

\begin{figure}
\centering
\includegraphics[width=0.9\columnwidth]{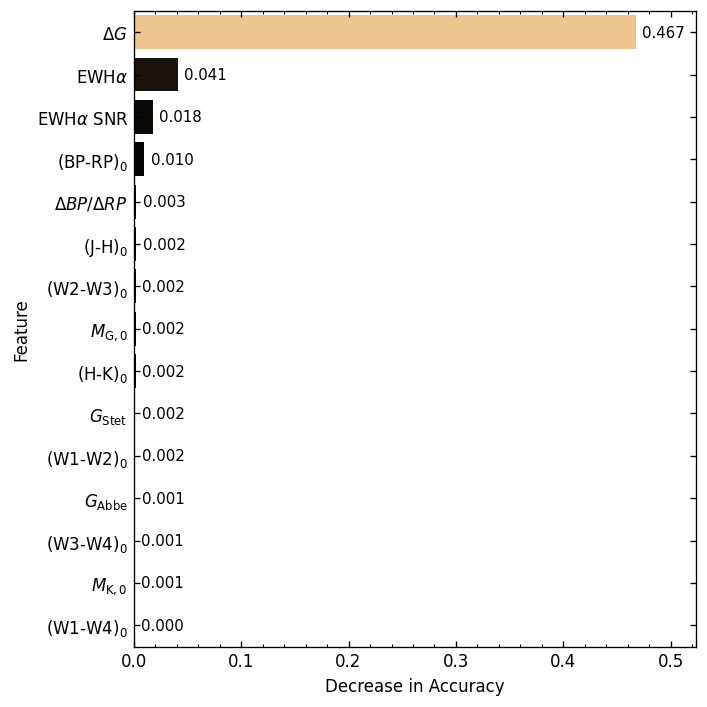}
\caption{Feature importance for the RF classifier. The ranking shows the relative contribution of individual \textit{Gaia} DR3 photometric, variability, and spectroscopic parameters to the classification, together with 2MASS and WISE colours.}
\label{fig:feature_importance}
\end{figure}

The relative importance of the input features was evaluated using permutation importance. The trimmed $G$-band variability amplitude is by far the most important parameter (Fig. \ref{fig:feature_importance}; as it allows to clearly separate Miras from the two other groups), followed by the pseudo-equivalent width of H$\alpha$, separating symbiotics from the contaminants The signal-to-noise ratio of the H$\alpha$ measurement also contributes to the classification, followed by the dereddened $(BP-RP)_0$ colour, while the remaining infrared colours and variability statistics have significantly smaller individual importance. This result is consistent with the comparisons presented above, which showed that variability amplitude, emission-line strength, and optical colour provide the strongest separation between symbiotic stars and pulsating red giants.

\subsubsection{Tests with reduced feature sets}

To evaluate the relative importance of different parameter groups, we constructed alternative versions of the classifier in which selected subsets of features were removed. In particular, we tested models without magnitudes and colours, without variability parameters, and without the H$\alpha$ measurements, while keeping the rest of the pipeline unchanged. The corresponding confusion matrices and feature-importance plots are shown in Appendix~\ref{app:fig}.

Interestingly, omitting the variability information affects the performance only slightly, and the classifier is still able to distinguish Miras, SR variables, and symbiotic stars, reaching a balanced accuracy of 0.926 (Fig.~\ref{fig:RF_no_variability}). In this case, the most important feature becomes the SNR ratio of the H$\alpha$ measurement (followed by the H$\alpha$ pEW on the third place). This is mainly because high H$\alpha$ SNR is typical for symbiotic stars, whereas both pulsating groups show low values on average, with SR variables having slightly higher values than Miras. The infrared colours then allow further separation of SRs and Miras, particularly for objects with dust excess.

If the absolute magnitudes and colour information are omitted (i.e. all external datasets, in practice including distance and extinction estimates), so that the classifier relies only on the H$\alpha$ measurements and variability parameters from \textit{Gaia} DR3, the performance becomes even slightly better, with a balanced accuracy of 0.940 (Fig.~\ref{fig:RF_no_colors}). In this configuration, the most important feature is the variability amplitude, while the H$\alpha$ measurements provide the main discrimination between symbiotic stars and single pulsating giants.

In the final experiment, we omitted the H$\alpha$ information. The separation between Miras and SR variables remains good (Fig. \ref{fig:RF_no_halpha}), with only a small fraction of these objects misclassified as symbiotic stars, likely because their colours differ from those of typical symbiotic systems, as discussed above. However, the performance for symbiotic stars themselves becomes poor: more than half of the known symbiotic stars are not classified as symbiotic in this version of the model, and this is not limited to objects with weak H$\alpha$ emission. In this configuration, the classifier identifies as symbiotic stars mainly those systems with particularly blue colours, while the majority of the population overlaps with normal pulsating giants in the remaining parameters.

\subsection{RF classification of DR3 sample}
\begin{figure}
\centering
\includegraphics[width=0.9\columnwidth]{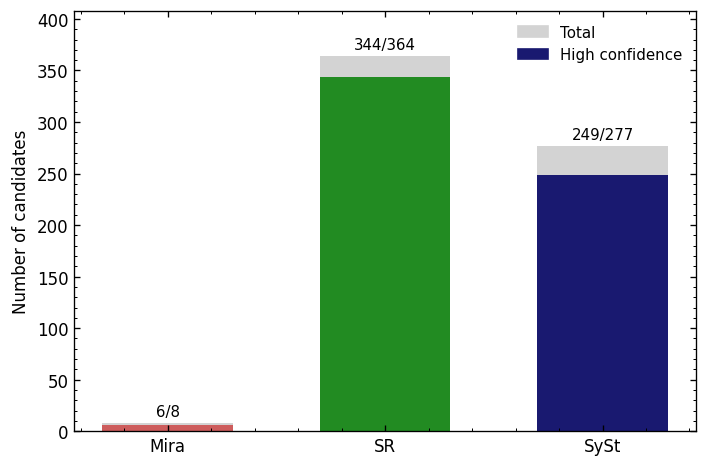}
\caption{Number of objects assigned to each class by the classifier for the full sample of 649 \textit{Gaia} DR3 candidates. The coloured portions of the columns indicate objects with classification probability higher than 0.75, which are considered high-confidence classifications, while the remaining objects correspond to lower-confidence assignments.}
\label{fig:candidate_numbers}
\end{figure}

We applied the classifier to the full sample of 649 \textit{Gaia} DR3 symbiotic star candidates. As described in Section~\ref{sec:sample}, this sample includes confirmed symbiotic stars, literature candidates, and new candidates selected from the \textit{Gaia} dataset. In total, 277 objects are classified as symbiotic stars by the pipeline, of which 249 have a classification probability higher than 0.75 and are therefore considered high-probability candidates (Fig.~\ref{fig:candidate_numbers}). 

Of the 246 confirmed symbiotic stars listed in NODSV, seven are not classified as symbiotic stars by the pipeline (namely M31SyS J004005.96+401604.3, M31SyS J004534.07+413049.0, V916 Sco, PPA J1746-3454, K 5-33, Sct X-1, and NGC 6822 SySt-1). These objects are either extragalactic systems or very faint sources for which the \textit{Gaia} DR3 parameters are incomplete. In most cases, the missing information concerns H$\alpha$ measurements (median imputation is used in such cases) or unreliable distance estimates, and therefore their misclassification is not unexpected and does not indicate a failure of the classifier.

The \textit{Gaia} DR3 sample also contains 61 previously proposed symbiotic candidates from the literature. Out of these, 27 are classified as symbiotic stars by the pipeline, while the remaining objects are assigned to the Mira or SR classes. {Some of these candidates may instead belong to other classes of red or emission-line objects that were not included in the training set, such as T Tauri stars or heavily reddened early-type emission-line stars. Since these classes were not represented during training, such objects would likely be classified as symbiotic stars because they share some of their photometric and spectroscopic properties, while differing from Mira and SR variables. We therefore do not attempt to further classify them in this work.} In addition, three objects listed in NODSV as misclassified symbiotic stars were included in the sample; one of them, Hen~3-1383, is classified as a symbiotic star by our pipeline, which is not surprising given that its colours are significantly bluer than those of normal pulsating giants (see Section~\ref{sec:sample}).

This leaves 10 objects among the 339 new candidates that are classified by the pipeline as likely symbiotic stars. Two of these objects, those with the lowest probability (0.51 and 0.57, respectively), lack H$\alpha$ measurements and were removed from the final candidate list because their absolute magnitudes and colours are inconsistent with those of symbiotic systems. The remaining eight objects represent the most promising new symbiotic star candidates identified by our selection from all \textit{Gaia} DR3 symbiotic candidates (see Table \ref{tab:rf_candidates}). 

\begin{table*}
\centering
\caption{Symbiotic candidates selected from our RF classification. Probabilities for each trained class are listed. If the object was classified as symbiotic star by \citet{2026MNRAS.546ag105A}, this is noted in the last column (see Section \ref{sec:akras}).}
\label{tab:rf_candidates}
\begin{tabular}{llrrrrrrl}
\hline
Name & \textit{Gaia} DR3 & RA [deg] & DEC [deg] & $G$ [mag] & $P_{\rm Mira}$ & $P_{\rm SR}$ & $P_{\rm SySt}$ & Note \\
\hline
2MASS J14141590-6718454 & 5847837327775225856 & 213.5662 & -67.3126 &  $15.01 \pm 0.01$ & 0.01 & 0.20 & 0.79 & \\
2MASS J17251717-2948346 & 4059074601049589760 & 261.3215 & -29.8096 & $14.93 \pm 0.01$ & 0.00 & 0.00 & 1.00 & in \citet{2026MNRAS.546ag105A} \\
V3664 Oph & 4110870983327739008 & 261.1665 & -24.3632 & $14.02 \pm 0.01$ & 0.06 & 0.01 & 0.92 & \\
PN Sa 3-119 & 4062924712807508736 & 271.6262 & -27.7399 & $13.22 \pm 0.01$ & 0.00 & 0.00 & 1.00 & in \citet{2026MNRAS.546ag105A}\\
Hen 3-1548 & 4043706834462070784 & 270.3422 & -32.2894 & $13.84 \pm 0.01$ & 0.00 & 0.00 & 1.00 & in \citet{2026MNRAS.546ag105A}\\
V469 Vul & 2025674084853852160 & 292.2177 & 27.1669  & $11.67 \pm 0.01$ & 0.00 & 0.00 & 1.00 & in \citet{2026MNRAS.546ag105A} \\
ATO J315.3668+45.9271 & 2163480206474053376 & 315.3668 & 45.9272  & $14.49 \pm 0.02$ & 0.05 & 0.00 & 0.95 & in \citet{2026MNRAS.546ag105A}\\
2MASS J21180196+5721343 & 2178988199495779456 & 319.5081 & 57.3596  & $12.78 \pm 0.01$ & 0.00 & 0.00 & 0.99 & \\
\hline
2MASS J20274687+3031193 & 1861714292419512960 & 306.9453 & 30.5220 & $14.17 \pm 0.00$ & 0.00 & 1.00 & 0.00 & H$\alpha$ imputed\\
\hline
\end{tabular}
\end{table*}

In addition, there are 21 objects classified by the pipeline as SR variables for which the H$\alpha$ pEW is not available in \textit{Gaia} DR3. For these objects, the H$\alpha$ value was imputed using the median of the training set, corresponding to a value close to zero, and therefore their resulting probability of belonging to the symbiotic class is negligible, as expected. In any case, most of these objects have luminosities inconsistent with symbiotic stars, with $M_G > 2.5$ mag and, when available, $M_K > 0.5$ mag. However, two objects show higher luminosities and cannot be safely rejected based on the available parameters, and therefore remain, in principle, possible symbiotic candidates. One of them, Gaia DR3 4052553741216415232, is in fact an incorrectly identified \textit{Gaia} counterpart of the previously considered symbiotic candidate PN M 1-44, which was included in the training set because it is listed as a candidate in \citet{2019ApJS..240...21A}. This object is, however, a superposition of the planetary nebula and a G–K type red giant \citep{2013MNRAS.432.3186M}. The nature of the second object (listed in Table \ref{tab:rf_candidates}) is discussed in Section~\ref{sec:our_candidates}.

The remaining objects among the 339 new candidates are most likely normal pulsating giants. A few sources are located close to the main sequence, and some may also belong to other types of contaminants, such as reddened hot variables that were not included in the training set. For simplicity, we adopted only two contaminant classes, pulsating giants and variable main-sequence stars, based on their position in the \textit{Gaia} CMD, using the separation between main-sequence and evolved stars defined by \citet{2025A&A...696A.243G} (see, e.g., their Fig.~6). Eleven objects lie in the vicinity of the main-sequence in the \textit{Gaia} CMD, while 319 are likely either SR or Mira pulsators.

We recommend moving all these objects to the "misclassified" group in NODSV, as there is no evidence supporting their symbiotic nature. In the following sections, we focus only on the eight strongest candidates selected by the full classifier, as well as an additional object without H$\alpha$ measurements but with a luminosity consistent with that of symbiotic stars. 

\section{Analysis of selected symbiotic candidates}\label{sec:our_candidates}
As described in the previous section, we selected a sample of eight strong symbiotic candidates from the 339 objects that were suggested for the first time as symbiotic candidates in \textit{Gaia} DR3 (and one without H$\alpha$ measurement). In this section, we present follow-up observations of these objects and discuss their nature. 

In addition to these candidates, we also observed eight further objects selected to serve as a control sample (see list in Table \ref{tab:obslog}). These targets were chosen to lie in relatively uncrowded regions for easy observations, to be bright ($G < 13.5$ mag), and to show no H$\alpha$ emission reported in \textit{Gaia} DR3 (pEW H$\alpha$ $> -10$\,\AA), that is, objects not classified as symbiotic candidates by our pipeline. This comparison sample allows us to assess how objects that do not satisfy our symbiotic selection criteria appear in spectroscopic follow-up.

\subsection{Observational data}
We conducted spectroscopic observations of the targets between July 2022 and May 2026 using a remotely controlled 0.3-m F/4 Newtonian telescope of the Southern Spectroscopic Project Observatory Team (2SPOT) consortium, located at Deep Sky Chile (DSC) near Cerro Tololo Observatory. Northern targets were observed with a 0.2-m F/5 Newtonian telescopes at Mont Ventoux and Cornillon observatories, a 0.35-m F/5 Schmidt-Cassegrain telescope at the Haute Provence Observatory (OHP), or a 1-m F/7 telescope at Calern Observatory. The telescopes were equipped with either a low-resolution Alpy600 spectrograph ($R \approx 600$) with a 23~$\mu$m slit or a LISA spectrograph with a 50~$\mu$m ($R \approx 500$) or 35~$\mu$m ($R \approx 750$) slit. Observations were carried out with ATIK 414EX and ATIK460 cooled cameras.   
This configuration provides a dispersion of approximately 2 --3~\AA\,pixel$^{-1}$ and covers a spectral range of 3800--7800~\AA\ for the Alpy600 and 4000--7500~\AA\ for the LISA spectrograph.

Individual exposures typically consisted of 600 or 1200~s subframes obtained with 1$\times$1 binning. Total integration times ranged from about 20~minutes to two hours, depending on the brightness of the target. Data reduction was performed using the {\tt Integrated Spectrographic Innovative Software} (ISIS)\footnote{\hyperlink{http://www.astrosurf.com/buil/isis-software.html}{http://www.astrosurf.com/buil/isis-software.html}}. All spectra were corrected for bias, dark, and flat-field frames using standard procedures. The instrumental response was derived from spectra of flux standard stars observed under similar conditions as the targets. Wavelength calibration was performed using an argon--neon lamp. No telluric correction was applied to the final spectra. The observing log is available in Table \ref{tab:obslog}.

\begin{figure*}
\centering
 \includegraphics[width=0.8\textwidth]{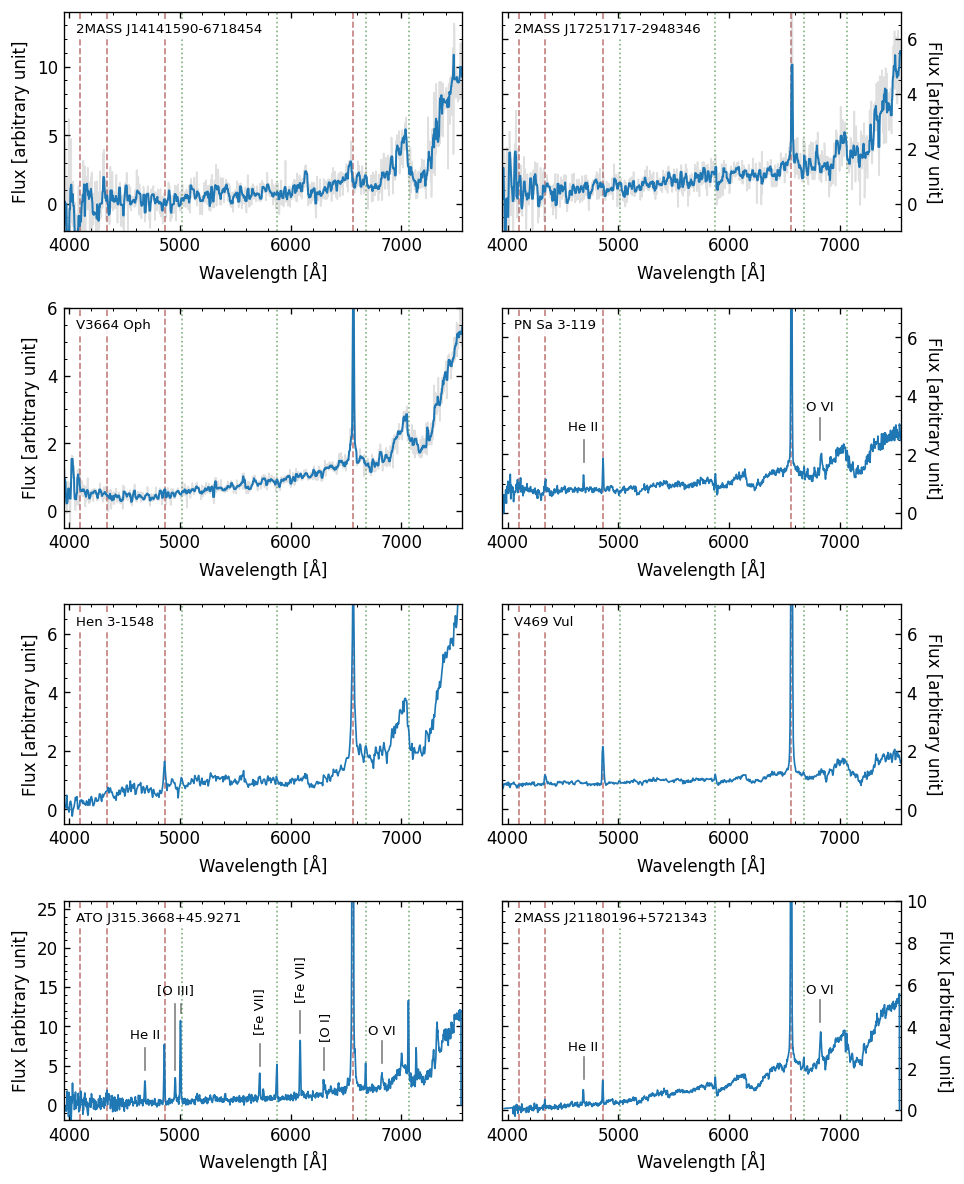}
 \caption{Follow-up spectra of symbiotic candidates identified by our RF pipeline. The positions of the Balmer lines are indicated with dashed red lines, while \ion{He}{i} lines are marked with dotted green lines. Additional emission lines of higher ionisation potential are shown in grey.}
 \label{fig:spectra}
 \end{figure*}

\subsection{Individual objects}
 
From the \textit{Gaia} DR3 sample, we have identified eight objects as very strong symbiotic candidates (Table~\ref{tab:rf_candidates}). Three of these show an M-type continuum with very strong emission lines, including Raman-scattered \ion{O}{vi}, and can therefore be unambiguously classified as newly confirmed symbiotic stars. Four additional objects exhibit only emission lines of low-ionisation potential, while one object shows no emission lines (Fig. \ref{fig:spectra}). Below, we discuss individual objects in more detail.\\

\noindent\textbf{2MASS J14141590-6718454}\\
This candidate received the lowest score among the eight objects in our classification pipeline and is also the reddest of all objects, which suggests negligible nebular contribution, if any. Its \textit{Gaia} light curve shows variability with a period of $\sim$331 days and an amplitude of 0.09 mag in the $G$ band \citep[][]{2023A&A...674A..15L}. While \textit{Gaia} DR3 suggests that the star is an emission-line object (with H$\alpha$ pEW = $-13.8 \pm 2.8$ \AA{}), our spectrum shows no emission lines (Fig. \ref{fig:spectra}). Therefore, there is no evidence supporting the symbiotic nature of this candidate, and we classify it as a pulsating giant.\\

\noindent\textbf{2MASS J17251717-2948346}\\
Our spectroscopy shows an M-type continuum with strong H$\alpha$ in emission (Fig. \ref{fig:spectra}). \citet[][]{2023A&A...674A..15L} reported a period of 147 days and an amplitude of 0.1 mag, but the \textit{Gaia} light curve also shows variability on a longer timescale of $\sim$600 days. Although the available data are not sufficient to confirm the symbiotic nature of this object, it remains a possible candidate.\\

\begin{figure}
\centering
\includegraphics[width=\columnwidth]{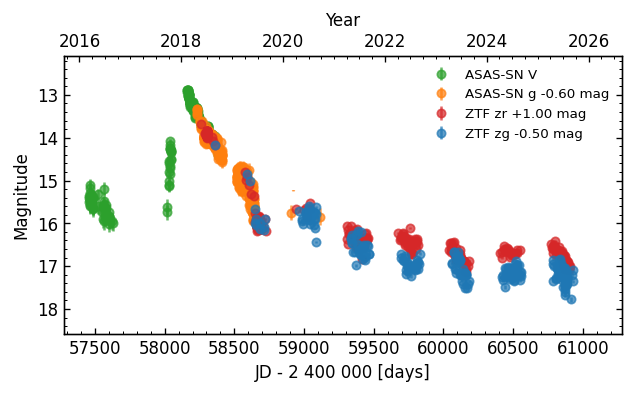}
\caption{ASAS-SN and ZTF light curves of V3664 Oph. The individual datasets were shifted by adding constant offsets to place them at the same level during the outburst for clarity.}
\label{fig:v3664_oph_lc}
\end{figure}

\noindent\textbf{V3664 Oph}\\
This star is classified as a Mira variable by \citet{2018AJ....156..241H} and is listed as such, e.g., in the SIMBAD database \citep[][]{2000A&AS..143....9W}. On the other hand, a brightening of the object was reported by T.~Kojima in 2018. The ASAS-SN light curve \citep[][]{2014ApJ...788...48S,2017PASP..129j4502K} confirms an outburst that started already in 2017 (Fig.~\ref{fig:v3664_oph_lc}). The pre-brightening level is not reliable due to blending with neighbouring sources. The decline from the brightening is also clearly visible in the ZTF photometry \citep[][]{2019PASP..131a8003M}. Spectroscopic follow-up obtained during the outburst with the SALT telescope classified the object as a classical nova \citep{2018ATel11338....1A}. The object was already known as an emission-line star before \citep[][]{2003AN....324..437K}.

The pre-eruption \textit{Gaia} XP spectrum shows clear TiO molecular bands and H$\alpha$ in emission, which is consistent with our spectroscopy from 2022 (JD 2\,459\,814.6) that also shows molecular bands and H$\alpha$ in emission (Fig. \ref{fig:spectra}). The $K$-band absolute magnitude of $\sim -6.5$ mag supports the presence of a red-giant donor in the system, which classifies the object as a symbiotic nova. We note that the possibility of a red-giant companion to this nova was already mentioned by \citet{2026MNRAS.546f2270C}. There is currently no evidence in the available photometry that the giant is pulsating as a Mira variable. It is possible that the outburst activity was mistaken for Mira variability, as, e.g., in the case of V390 Sco \citep{2024AN....34540017M}.\\ 

\noindent\textbf{PN Sa 3-119}\\
This object was originally considered to be a planetary nebula \citep[][]{1976PW&SO...2...57S}, but was later reclassified as an "M star" \citep[][]{2001A&A...378..843K}. Our spectrum confirms an M-type continuum and shows strong Balmer lines in emission together with \ion{He}{i}, \ion{He}{ii}, and Raman-scattered \ion{O}{vi} (Fig. \ref{fig:spectra}), which undoubtedly classifies the star as a symbiotic system. \citet{2023A&A...674A..15L} reported a period of 786 days in the \textit{Gaia} light curve with an amplitude of $\sim$0.1 mag in the $G$ band, which is likely the orbital period of this system.\\

\noindent\textbf{Hen 3-1548}\\
This candidate was known as an H$\alpha$ emitter, and \citet{1976ApJS...30..491H} and \citet{1977PASP...89..901C} mentioned the possibility that it could be a nova because of reported brightness changes. Our spectroscopy shows an M-type continuum with Balmer lines in emission (Fig. \ref{fig:spectra}). The available photometry does not show any clear outburst activity, but we keep this object as a possible symbiotic candidate.\\

\noindent\textbf{V469 Vul}\\
This star, also known as AS~357, is a long-known emission-line object \citep[][]{1950ApJ...112...72M} and has also been considered an OB star \citep{2003AJ....125.2531R}. \citet{2023A&A...674A..15L} reported a period of 788 days in the \textit{Gaia} light curve with an amplitude of $\sim$0.13 mag in the $G$ band. Our observations reveal an M-type continuum with emission lines of \ion{H}{i} and \ion{He}{i} (Fig. \ref{fig:spectra}). While the lack of higher-ionisation emission lines prevents us from classifying it as a genuine symbiotic star at the moment, it remains a strong candidate, probably being an accreting-only symbiotic system. The presence of molecular bands rules out the possibility of a reddened hot star.\\

\noindent\textbf{ATO J315.3668+45.9271}\\
This is the second clear case in the sample. Our spectrum shows an M-type continuum with strong emission lines of \ion{H}{i}, \ion{He}{i}, \ion{He}{ii}, [\ion{O}{iii}], [\ion{Fe}{vii}], and Raman-scattered \ion{O}{vi} (Fig. \ref{fig:spectra}). We note that we independently identified this object as a symbiotic star in an analysis of \textit{Gaia} planetary nebula candidates selected from the emission-line star pipeline of \textit{Gaia} DR3, and provide a more detailed analysis in that work \citep[][]{2026arXiv260424730M}. The symbiotic nature of this object was also supported by the independent analysis of \citet{2025AstBu..80..620T}.\\

\noindent\textbf{2MASS J21180196+5721343}\\
Similarly to the previous candidate, this object shows an M-type continuum with emission lines of \ion{H}{i}, \ion{He}{i}, \ion{He}{ii}, and rather strong \ion{O}{vi} (Fig. \ref{fig:spectra}), confirming its symbiotic nature. Moreover, as we reported in \citet{2024A&A...682A...7B}, this object is included among the SB1 binaries in \textit{Gaia} DR3 \citep{2023A&A...674A...1G}, with an orbital period of 754 days, eccentricity of 0.28, and radial-velocity semi-amplitude of 10.3 km\,s$^{-1}$.\\

\noindent\textbf{2MASS J20274687+3031193}\\
The probability of this object being a symbiotic star in our full RF pipeline is negligible. However, this source does not have an H$\alpha$ pEW value available in the \textit{Gaia} DR3 table, and therefore the missing value was replaced in the pipeline by the median of the training set, i.e., a value close to zero. It is therefore not surprising that the object was not classified as a symbiotic star. Since its absolute magnitudes fall within the range typical for symbiotic stars, we obtained spectroscopic follow-up observations to allow a reliable classification. As the spectrum shows no emission lines (Fig. \ref{fig:spectra_check}), we classify this object as a pulsating giant rather than a symbiotic system.

{To summarise, of the nine objects discussed, three (PN Sa 3-119, ATO J315.3668+45.9271, and 2MASS J21180196+5721343) exhibit highly ionised emission lines and are unambiguously confirmed as new symbiotic stars. Three additional objects (2MASS J17251717-2948346, Hen 3-1548, and V469 Vul) show only low-ionisation emission lines and are therefore classified as possible symbiotic stars. V3664 Oph likewise displays only low-ionisation emission lines; however, in combination with its 2017–2019 nova outburst, we classify it as a symbiotic nova. Finally, 2MASS J14141590-6718454 and 2MASS J20274687+3031193 are classified as pulsating giants.}

To complete the discussion, as mentioned at the beginning, we also observed an additional eight stars from the \textit{Gaia} DR3 sample, all of which were safely rejected as symbiotic by our pipeline. Their spectra are shown in Fig.~\ref{fig:spectra_control}. Four of the objects are M-type giants, two are clearly C-rich giants, and one additional object, although observed with a low S/N spectrum, is a known C-rich star in the literature. The remaining object is a hot star, confirming that the \textit{Gaia} DR3 symbiotic sample is affected by some contamination from reddened hot stars. As expected, none of the spectra display emission lines.

\section{Assessment of the symbiotic candidates proposed by Akras et al. (2026)}\label{sec:akras}

{Recently, \citet{2026MNRAS.546ag105A} conducted an independent and, in some respects, similar analysis of the same \textit{Gaia} DR3 sample of 649 symbiotic candidates. However, their methodology differs substantially from the one adopted in this work. While we employ a supervised RF classifier that simultaneously combines photometric, spectroscopic, and variability-related parameters to distinguish symbiotic stars from their dominant contaminants, \citet{2026MNRAS.546ag105A} applied a sequence of empirical photometric selection criteria to identify the most promising candidates, without discussing the nature of the remaining objects. They primarily used photometry from IPHAS, VPHAS+, 2MASS, and WISE for this purpose. They also incorporated H$\alpha$ information in their analysis, but in a different way from our approach. In particular, they either used H$\alpha$ measurements from IPHAS/VPHAS+ or computed synthetic H$\alpha$ magnitudes in these filters from the \textit{Gaia} XP spectra.}

In their study, the \textit{Gaia} sample was cross-matched with their catalogue \citep[][]{2019ApJS..240...21A} and NODSV, resulting in 246 confirmed symbiotic stars. The candidate symbiotic stars from the literature were not filtered at this stage. For the remaining objects, they applied a series of colour-based selection criteria and obtained a list of 13 promising candidates (10 S-type and 3 D-type). They concluded that, although spectroscopic follow-up is typically required to confirm symbiotic stars, the availability of \textit{Gaia} XP spectra allows them to classify all 13 objects as bona-fide symbiotic systems. While we agree that these objects represent very strong candidates, we caution against classifying them as bona-fide symbiotic stars without optical spectroscopy. We therefore discuss these 13 sources individually in this section.

\begin{figure*}
\centering
 \includegraphics[width=0.8\textwidth]{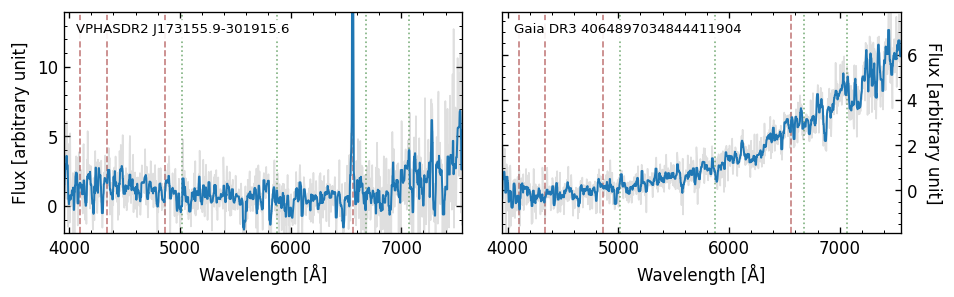}
 \caption{Spectra of two candidates classified as symbiotic stars by \citet{2026MNRAS.546ag105A}, for which no spectra were previously available in the literature. The positions of the Balmer lines are indicated with dashed red lines, while \ion{He}{i} lines are marked with dotted green lines.}
 \label{fig:spectra_akras}
 \end{figure*}

\citet{2026MNRAS.546ag105A} have five objects (marked in Table \ref{tab:rf_candidates}) in common with the sample of our eight candidates discussed in Section~\ref{sec:our_candidates}. As noted there, two of these are clear symbiotic stars, and one is likely an accreting-only system, while the remaining two show an M-type continuum with only H$\alpha$ emission and therefore cannot be conclusively classified as symbiotic stars.

An additional six candidates listed by \citet{2026MNRAS.546ag105A} were already known as symbiotic candidates and have optical spectra available in the literature. 
[MMU2013] 001.71+01.14 and [MMU2013] 001.33+01.07 were classified as possible symbiotic stars by \citet{2013MNRAS.432.3186M}. Their SALT spectra show TiO bands together with \ion{H}{i}, \ion{He}{i}, and \ion{Ca}{ii} triplet emission, however, they were not considered bona-fide symbiotic systems because they do not fulfill all commonly adopted criteria \citep[see, e.g.,][]{2000A&AS..146..407B,2013MNRAS.432.3186M,2021MNRAS.506.4151M,2026MNRAS.545S2146M}, in particular the presence of highly ionized emission lines. 
For [MMU2013] 357.12+01.66, the authors noted similarities with D-type symbiotic stars, but the spectrum shows only \ion{H}{i} and \ion{O}{i} emission, which is not sufficient for a secure classification.

WRAY 16-294 was already suspected to be of symbiotic nature by \citet{1997A&A...327..191M} (note that it was mistyped as WRAY 16-296 in that work), who reported a red continuum consistent with a reddened K-type giant together with emission lines of \ion{H}{i}, \ion{He}{i}, and faint [\ion{O}{iii}]. When combined with the \textit{Gaia} data and the analysis by \citet{2026MNRAS.546ag105A}, the symbiotic classification of this object appears very likely.

[KW2003] 55 was classified as a symbiotic candidate by \citet{2014MNRAS.440.1410M}, who noted that its spectrum resembles that of an S-type symbiotic system. However, similarly to the cases above, the highly ionised emission lines required to confirm the presence of a hot companion are missing, and the symbiotic nature could not be firmly established. 
\citet{2026MNRAS.545S2146M} recently suggested that this may be due to the object being observed during a prolonged outburst, but only additional spectroscopy obtained during quiescence can test this hypothesis.

IRAS 17114-3929 was suggested as a D-type candidate based on its near-IR colours and PN-like optical spectrum by \citet{2014MNRAS.440.1410M}, however, the presence of the red giant component has not yet been confirmed, and an alternative classification as a planetary nebula with a dusty Wolf-Rayet central star cannot be excluded.

This leaves two additional objects.  
VPHASDR2 J173155.9-301915.6 was already proposed as a symbiotic star by \citet{2019MNRAS.483.5077A}, but no spectroscopic confirmation appears to be available in the literature. Our pipeline also classifies this object as a symbiotic star. Its \textit{Gaia} XP spectrum shows clear detection of H$\alpha$ in emission and molecular bands in the red part of the spectrum. We have obtained a follow-up spectrum in the same way as for the rest of the objects in this study. Although it has low S/N, we can clearly detect strong H$\alpha$ and a hint of molecular bands (Fig. \ref{fig:spectra_akras}). However, the spectrum is not sufficient for the definitive classification of the object as a symbiotic star, as we cannot detect other emission lines in the spectrum.

Gaia DR3 4064897034844411904 (=2MASS J18092210-2538513) was not previously suspected to be of symbiotic nature prior to \textit{Gaia} DR3. In our pipeline, it is classified as an SR variable with a probability of 0.990, and it does not show significant H$\alpha$ emission in \textit{Gaia} DR3 (pEW H$\alpha$ = $0.0 \pm 5.8$ \AA{}). It is therefore unclear why \citet{2026MNRAS.546ag105A} classified this object as a symbiotic binary. Our follow-up spectrum shows no emission in the H$\alpha$ region (Fig. \ref{fig:spectra_akras}), and the object, therefore, is unlikely to be a symbiotic system.

In summary, of the 13 candidates proposed by \citet{2026MNRAS.546ag105A}, only two can be securely confirmed as symbiotic stars with the currently available data. The remaining sources, except the last one, remain promising candidates, but their symbiotic nature cannot be considered established without additional observations.

\section{Conclusions}\label{sec:conclusions}

We analysed the sample of 649 objects classified as symbiotic stars in the general variability classification of \textit{Gaia} DR3 in order to assess the reliability of this class and to identify new genuine symbiotic systems. Our main conclusions can be summarised as follows:

\begin{itemize}
\item Inspection of the training set used in the \textit{Gaia} DR3 variability classification shows that it contains a mixture of confirmed symbiotic stars, candidates, and a few incorrectly identified objects. This may have affected the resulting symbiotic sample.

\item Of the 649 objects classified as symbiotic stars in \textit{Gaia} DR3, 246 correspond to previously known systems, and 61 were already listed as candidates in the literature, while 339 objects represent new candidates proposed for the first time by the \textit{Gaia} variability pipeline. Simple diagnostics based on the \textit{Gaia} colour-magnitude diagram, near-infrared colours, and the \textit{Gaia} pseudo-equivalent width of H$\alpha$ indicate that a large fraction of these new candidates are likely contaminants, primarily pulsating red giants.

\item To quantify the level of contamination, we constructed a Random Forest classifier trained on confirmed symbiotic stars and on Mira and semi-regular variables, which represent the dominant expected source of false positives. Using \textit{Gaia} photometry, variability parameters, H$\alpha$ measurements from XP spectra, and infrared colours, the classifier reaches a balanced accuracy of $\approx$0.94 in cross-validation and successfully separates most symbiotic systems from single pulsating giants.

\item Among the 339 new candidates, only eight objects remain strong symbiotic candidates after the full selection using our classifier. This confirms that the majority of sources classified as symbiotic in the \textit{Gaia} DR3 variability pipeline are not genuine interacting binaries.

\item We obtained follow-up spectroscopy for these candidates and for a comparison sample of objects rejected by the classifier. Three of the selected candidates (PN Sa 3-119, ATO J315.3668+45.9271, and 2MASS J21180196+5721343) show clear high-excitation emission lines, including Raman-scattered O VI, and can be unambiguously confirmed as new symbiotic stars. Three additional objects (2MASS J17251717-2948346, Hen 3-1548, and V469 Vul) display only emission lines of low ionisation and remain possible symbiotic systems. V3664 Oph also shows only low-ionisation; however, combined with the 2017-2019 nova outburst, we classify it as a symbiotic nova. The control sample confirms that objects rejected by the classifier do not show emission-line spectra, supporting the validity of our selection procedure.

\item Our results suggest that the purity of the \textit{Gaia} DR3 variability classification is low due to strong overlap with pulsating red giants in the parameter space used by the classifier. The inclusion of spectroscopic indicators, in particular the H$\alpha$ pseudo-equivalent width from XP spectra, is essential for reliable separation of symbiotic binaries from single evolved stars.

\item The small number of new confirmed systems found among the \textit{Gaia} DR3 candidates implies that the discrepancy between the predicted and observed number of symbiotic stars in the Galaxy remains unresolved. Nevertheless, \textit{Gaia} provides a powerful starting point for systematic searches, especially when combined with additional photometric, spectroscopic, and variability diagnostics.
\end{itemize}

Future Gaia data releases, together with multi-wavelength surveys and targeted spectroscopic follow-up, will be crucial for constructing a more complete and homogeneous census of symbiotic binaries in the Milky Way.

\section*{Acknowledgements}
{We thank the referee for a careful reading of the manuscript and for the constructive comments and suggestions, which helped improve the quality and clarity of this paper.} We would like to thank Laurent Bernasconi for hosting our equipment in his domes at Mont Ventoux. The research of JMe was supported by the Czech Science Foundation (GACR) project no. 24-10608O. JMe and PGB acknowledge support by the Spanish Ministry of Science and Innovation with the grant no. PID2023-146453NB-100 (PLAtoSOnG). JMik was supported by the Polish National Science Centre (NCN) grant 2023/48/Q/ST9/00138. PGB was supported by the Spanish Ministry of Science, Innovation, \& Universities (MCIN) with the Ramón y Cajal fellowship (RYC-2021-033137-I). AE and MAM received the support of two fellowships from “La Caixa” Foundation (ID 100010434). The fellowship codes are LCF/BQ/PI23/11970031 (PI: Ana Escorza) and LCF/BQ/PI23/11970035 (PI: Michael Abdul-Masih). This work has made use of data from the European Space Agency (ESA) mission
{\it Gaia} (\url{https://www.cosmos.esa.int/gaia}), processed by the {\it Gaia}
Data Processing and Analysis Consortium (DPAC,
\url{https://www.cosmos.esa.int/web/gaia/dpac/consortium}). Funding for the DPAC
has been provided by national institutions, in particular the institutions
participating in the {\it Gaia} Multilateral Agreement. 

\section*{Data Availability}
The spectroscopic data underlying this article will be shared on reasonable request to the corresponding author. Additional data used in this work are publicly available from the archives of the ASAS-SN, ZTF, and \textit{Gaia} surveys.



\bibliographystyle{mnras}
\bibliography{lit} 

\bsp	
\label{lastpage}

\newpage
\FloatBarrier
\appendix
\section{\textit{Gaia} DR3 counterpart of AS 269}\label{app:as269}
During our analysis, we noticed an interesting case involving the confirmed symbiotic star AS~269. Two sources classified as SYST in \textit{Gaia} DR3 (\textit{Gaia} DR3 4042885769793130752 and \textit{Gaia} DR3 4042885774159741952), separated by only 0.58 arcsec, lie very close to the catalog position of AS~269 (at distances of 0.23 and 0.35 arcsec, respectively).

AS~269 was initially considered a possible planetary nebula by \citet{1967ApJS...14..125H} and later classified as a B[e] star by \citet{1973ApJ...185..899S}. Subsequent studies by \citet{1985MNRAS.213...59W} and \citet{1997A&A...327..191M} identified it as a suspected symbiotic star. \citet{1997A&A...327..191M} noted that the object shows emission-line spectra resembling those of Be stars, while the continuum slope suggests a G-type star and the near-IR colours are consistent with a reddened G5–K2 giant. They also pointed out similarities with the yellow D$'$-type symbiotic system HD~330036, which contains a G-type giant surrounded by a warm dust shell.

In current catalogs (e.g., NODSV), the symbiotic star is associated with the first of the above \textit{Gaia} sources, likely because it lies slightly closer to the catalog position. Only a single 2MASS and WISE source is present, located approximately in the middle between the two \textit{Gaia} detections. A similar issue of incorrect \textit{Gaia} counterpart identification was recently reported for another symbiotic star, Terz~V~2513 \citep{2026MNRAS.545f2094M}, where the catalog position was matched to a nearby but unrelated Mira variable.

In the present case, however, the situation appears different. Both \textit{Gaia} sources have nearly identical photometry: $G = 13.90 \pm 0.02$ and $13.87 \pm 0.01$, $BP = 13.46 \pm 0.02$ and $13.46 \pm 0.01$, and $RP = 12.39 \pm 0.02$ and $12.43 \pm 0.01$. Only the second source (\textit{Gaia} DR3 4042885774159741952) has astrometric parameters and low-resolution XP spectra available (which confirms the presence of emission lines). The first source has significantly fewer field-of-view (FoV) transits (e.g., 11 used for astrometry compared to 41 for the second source). Similarly, the numbers of photometric observations are much lower: 130 and 350 measurements in $G$, 10 and 34 in $BP$, and 11 and 34 in $RP$ for the first and second source, respectively. 

For comparison, sources of similar brightness in this field typically have more than 50 astrometric FoV transits, more than 450 $G$ observations, and roughly 50 measurements in $BP$ and $RP$. We therefore suspect that both entries correspond to the same physical source that was split during the \textit{Gaia} data processing (see \citealt{2021A&A...649A..10T}). In this case, \textit{Gaia} DR3 4042885774159741952 should be considered the correct counterpart of AS~269.

\section{Misclassified training set SRs and Mira} \label{app:misclassified}

\begin{figure*}
\centering
 \includegraphics[width=0.8\textwidth]{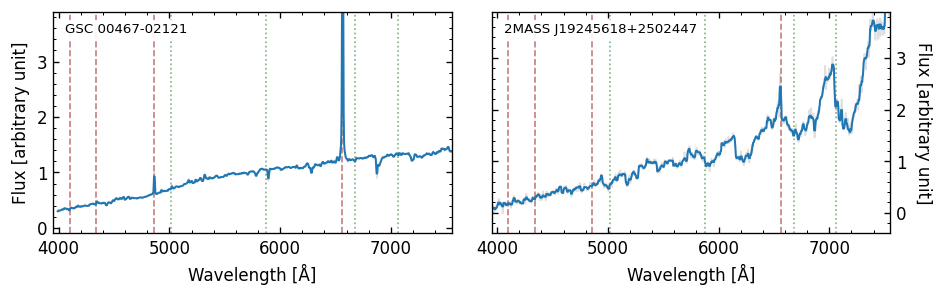}
 \caption{Spectra of two bright objects from the SR training set that were classified as symbiotic stars by our RF classifier. The positions of the Balmer lines are indicated with dashed red lines, while \ion{He}{i} lines are marked with dotted green lines.}
 \label{fig:spectra_training_set}
 \end{figure*}

\begin{table*}
\centering
\caption{Objects from the training set with high SySt probability but different true class.}
\label{tab:false_syst_candidates}
\begin{tabular}{llcrrcccc}
\hline
Name & \textit{Gaia} DR3 & RA [deg] & DEC [deg] & $G$ [mag] & $P_{\rm Mira}$ & $P_{\rm SR}$ & $P_{\rm SySt}$ & Training class \\
\hline
GSC 00467-02121        & 4268104444583376768 & 287.3201 &  2.5087  & 10.64 $\pm$ 0.00 & 0.0068 & 0.0096 & 0.9836 & SR   \\
ZTF J191508.94+120439.2 & 4312986474825074688 & 288.7873 & 12.0776  & 17.52 $\pm$ 0.00 & 0.0051 & 0.2565 & 0.7384 & SR   \\
OGLE BLG-LPV-234169     & 4112359240978047232 & 257.5534 & -25.6673 & 15.40 $\pm$ 0.04 & 0.3630 & 0.0052 & 0.6318 & Mira \\
ZTF J191103.00+052638.0 & 4293573634983790976 & 287.7625 &  5.4439  & 16.82 $\pm$ 0.01 & 0.0034 & 0.4224 & 0.5742 & SR   \\
2MASS J19245618+2502447 & 2023118300119332608 & 291.2341 & 25.0458  & 12.73 $\pm$ 0.00 & 0.0085 & 0.4569 & 0.5347 & SR   \\
ZTF J185758.98+001645.1 & 4265881300782664064 & 284.4958 &  0.2792  & 17.04 $\pm$ 0.01 & 0.0106 & 0.4913 & 0.4981 & SR   \\
\hline
\end{tabular}
\end{table*}

In our full RF classifier (Section \ref{sec:rf}), five SR variables and one Mira from the training set were classified as symbiotic stars in the classifier (Table \ref{tab:false_syst_candidates}). We briefly investigate these objects here. For two relatively bright objects, we also obtained the optical follow-up with the same setup as for symbiotic candidates analyzed in this work.\\ 

\noindent\textbf{GSC 00467-02121}\\
This star is a known H$\alpha$ emitter \citep[][]{1997AAHam..11.....K,1999A&AS..134..255K} and has previously been classified as a Be or OBe star \citep[][]{1955BOTT....2m..19I,1956BOTT....2n..31I,1977PW&SO...2...71S}.
While its \textit{Gaia} DR3 H$\alpha$ pEW of $-21.6 \pm 0.7$ \AA{}, together with its absolute magnitudes 
($M_{G,0} = -2.44$ mag; $M_{K,0} = -4.59$ mag) and colors 
(e.g., $(BP-RP)_0 = 1.19$ mag; $(J-H)_0 = 0.50$ mag; $(W1-W2)_0 = 0.44$ mag), make it a plausible symbiotic candidate, the source does not lie near the center of the symbiotic-star population in the parameter space considered here. It is therefore possible that the object represents another type of emission-line source that has been misclassified in the literature as an LPV and is not represented in our algorithm. Its \textit{Gaia} XP spectrum does not show clear molecular bands. We obtained an optical spectrum of the source (Fig.~\ref{fig:spectra_training_set}), which reveals prominent Balmer emission lines together with He~I and Fe~II emission, but no higher-ionization lines or molecular features. The extinction toward the source is non-negligible (total extinction $E(B-V) = 0.88$ mag; \citealt{2011ApJ...737..103S}), suggesting that the object is most likely indeed a reddened classical Be/OBe star. \\

\noindent\textbf{ZTF J191508.94+120439.2}\\ 
This object is classified as a SR variable with a period of 39.8 days in the ZTF $r$ band 
\citep[][]{2020ApJS..249...18C}. In \textit{Gaia} DR3 it shows strong H$\alpha$ emission 
(pEW H$\alpha$ = $-27.8 \pm 3.2$ \AA{}). However, its absolute magnitudes and intrinsic colors are uncertain because the 
extinction in its direction is very high (total extinction $E(B-V) = 6.99$ mag; \citealt{2011ApJ...737..103S}), 
making the derived dereddened parameters strongly distance dependent.\\

\noindent\textbf{OGLE BLG-LPV-234169}\\ 
This target is a known Mira variable with a period of 342.3 days and an amplitude of 3.0 mag in the $I$ band 
\citep[][]{2022ApJS..260...46I}. The reported \textit{Gaia} DR3 H$\alpha$ pEW of 
$-45.8 \pm 7.3$ \AA{} makes it a viable candidate for a symbiotic Mira.\\

\noindent\textbf{ZTF J191103.00+052638.0}\\
Another object that is classified as an SR variable. It has a period of 55.5 days in the ZTF $r$ band 
\citep[][]{2020ApJS..249...18C}. A strong H$\alpha$ emission line is reported in \textit{Gaia} DR3 
(pEW H$\alpha$ = $-24.6 \pm 7.1$ \AA{}). Its colors and absolute magnitudes place it within the region populated by known 
symbiotic stars, although in our classification pipeline, the probability of being an SR variable is only slightly lower 
than that of being a symbiotic star.\\

\noindent\textbf{2MASS J19245618+2502447}\\
This star is classified as an LPV with a photometric period of 205.5 days 
(amplitude 0.08 mag in the $G$ band) and a radial-velocity period of 448.9 days with an amplitude of 8.74 km\,s$^{-1}$ 
in the \textit{Gaia} FPR \citep[][]{2023A&A...680A..36G}. 
The H$\alpha$ emission reported in \textit{Gaia} DR3 is weaker (pEW H$\alpha$ = $-9.7 \pm 1.0$ \AA{}), 
however its colors and absolute magnitude place it within the bulk of the symbiotic population. 
In our pipeline, the probability of it being an SR variable is only slightly lower than the probability of it being a symbiotic star. Our spectroscopy reveals a clear M-type continuum together with H$\alpha$ emission (right panel of Fig. \ref{fig:spectra_training_set}). While these characteristics alone are insufficient for a definitive symbiotic classification, the possible detection of orbital variability, likely including an ellipsoidal modulation, supports the interpretation of the object as a candidate accretion-powered symbiotic star.\\

\noindent\textbf{ZTF J185758.98+001645.1}\\
This object is classified as an SR variable with a period of 49.5 days in the ZTF $r$ band 
\citep[][]{2020ApJS..249...18C}. According to \textit{Gaia} DR3, it shows strong H$\alpha$ emission 
(pEW H$\alpha$ = $-38.5 \pm 14.6$ \AA{}), although the SNR ratio of the measurement is relatively low, 
making the strength of the emission uncertain. No XP spectrum is available for this object. In our pipeline, the object received nearly identical probabilities of being classified as a symbiotic star and as an SR variable. 
\clearpage

\begin{figure}
\section{Additional figures and tables}\label{app:fig}
\centering
\includegraphics[width=\columnwidth]{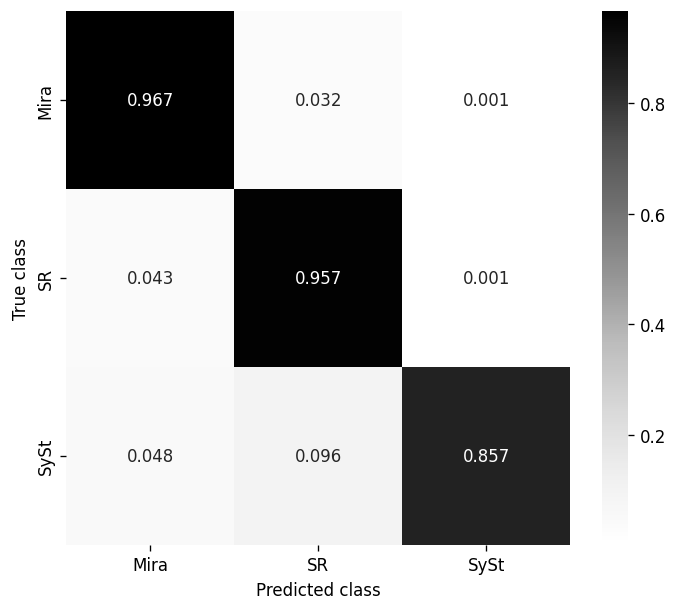}
\includegraphics[width=\columnwidth]{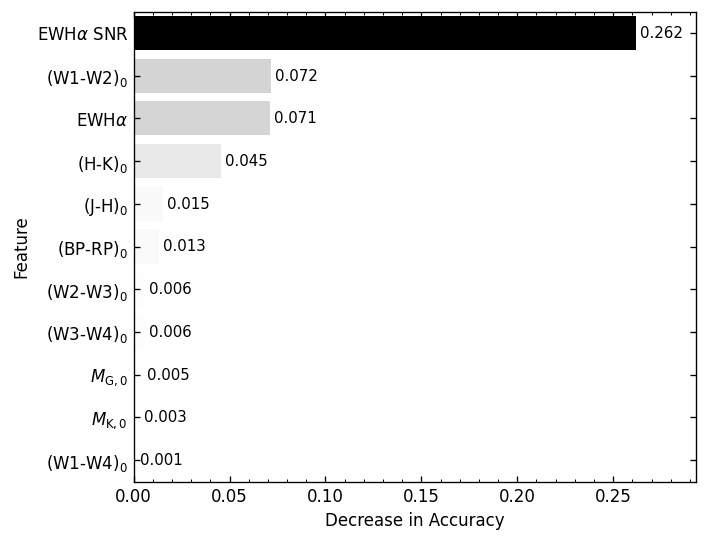}

\caption{Confusion matrix (top) and feature importance (bottom) for the classifier constructed without variability parameters. The performance is only slightly degraded compared to the full model, showing that the H$\alpha$ measurements and infrared colors provide strong discrimination between symbiotic stars, SR variables, and Miras.}
\label{fig:RF_no_variability}
\end{figure}

\begin{figure}
\centering
\includegraphics[width=\columnwidth]{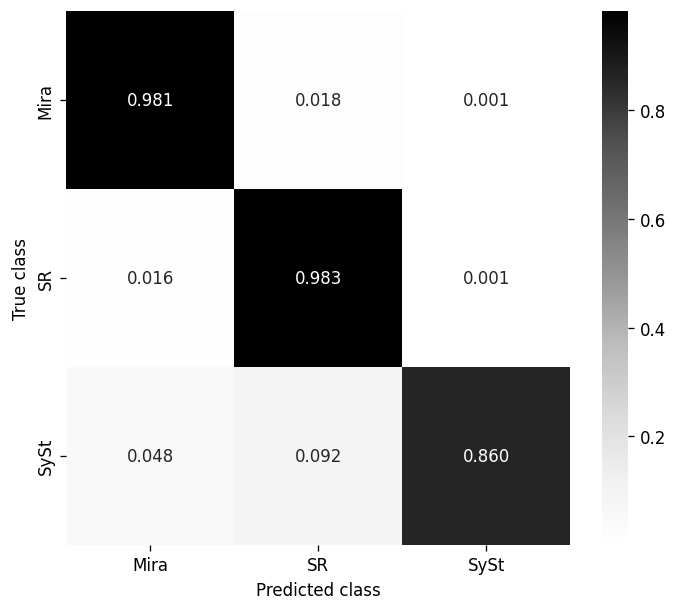}
\includegraphics[width=\columnwidth]{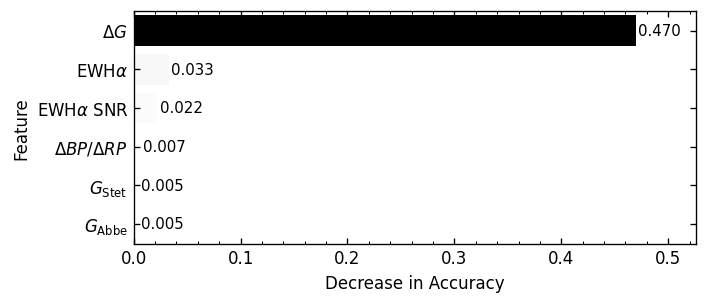}

\caption{Same as Fig. \ref{fig:RF_no_variability} for the classifier constructed without absolute magnitudes and color information, i.e. without using external datasets such as distance and extinction estimates. The classifier relies only on variability and H$\alpha$ parameters and achieves a performance almost comparable to the full model.}
\label{fig:RF_no_colors}
\end{figure}

\begin{figure}
\centering
\includegraphics[width=\columnwidth]{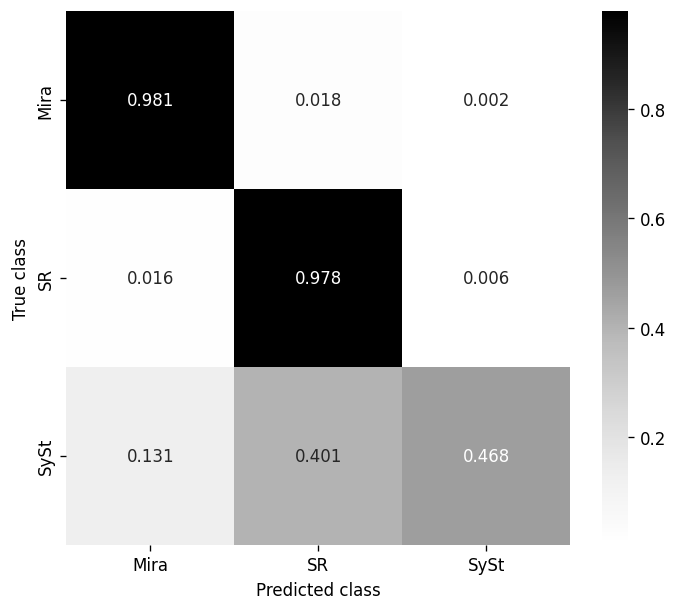}
\includegraphics[width=\columnwidth]{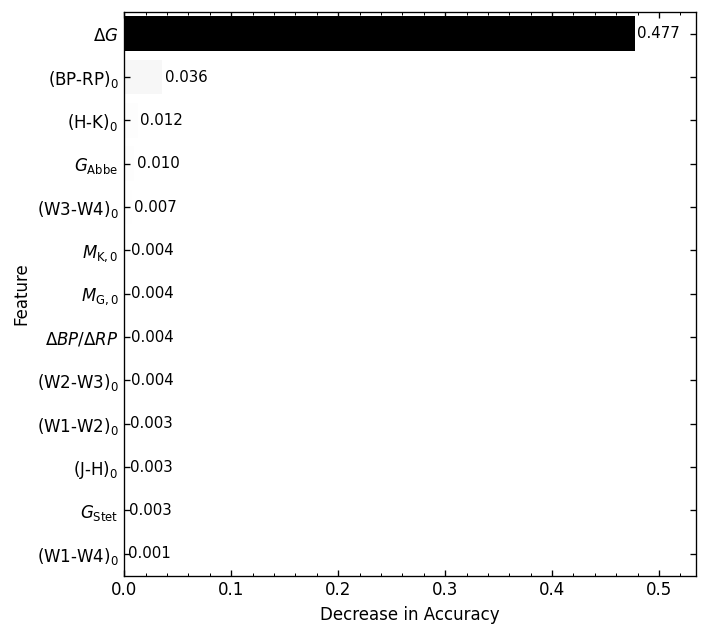}

\caption{Same as Fig. \ref{fig:RF_no_variability} for the classifier constructed without H$\alpha$ measurements. While the separation between Miras and SR variables remains good, the performance for symbiotic stars is significantly degraded, with a large fraction of known symbiotic stars misclassified as pulsating giants.}
\label{fig:RF_no_halpha}
\end{figure}

\begin{figure}

\centering
\includegraphics[width=\columnwidth]{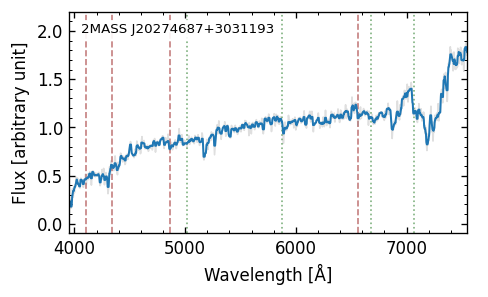}

\caption{Spectrum of 2MASS J20274687+3031193. Expected positions of the Balmer lines are indicated with dashed red lines, while those of \ion{He}{i} lines are marked with dotted green lines.}
\label{fig:spectra_check}
\end{figure}

\begin{figure*}
\centering
 \includegraphics[width=0.8\textwidth]{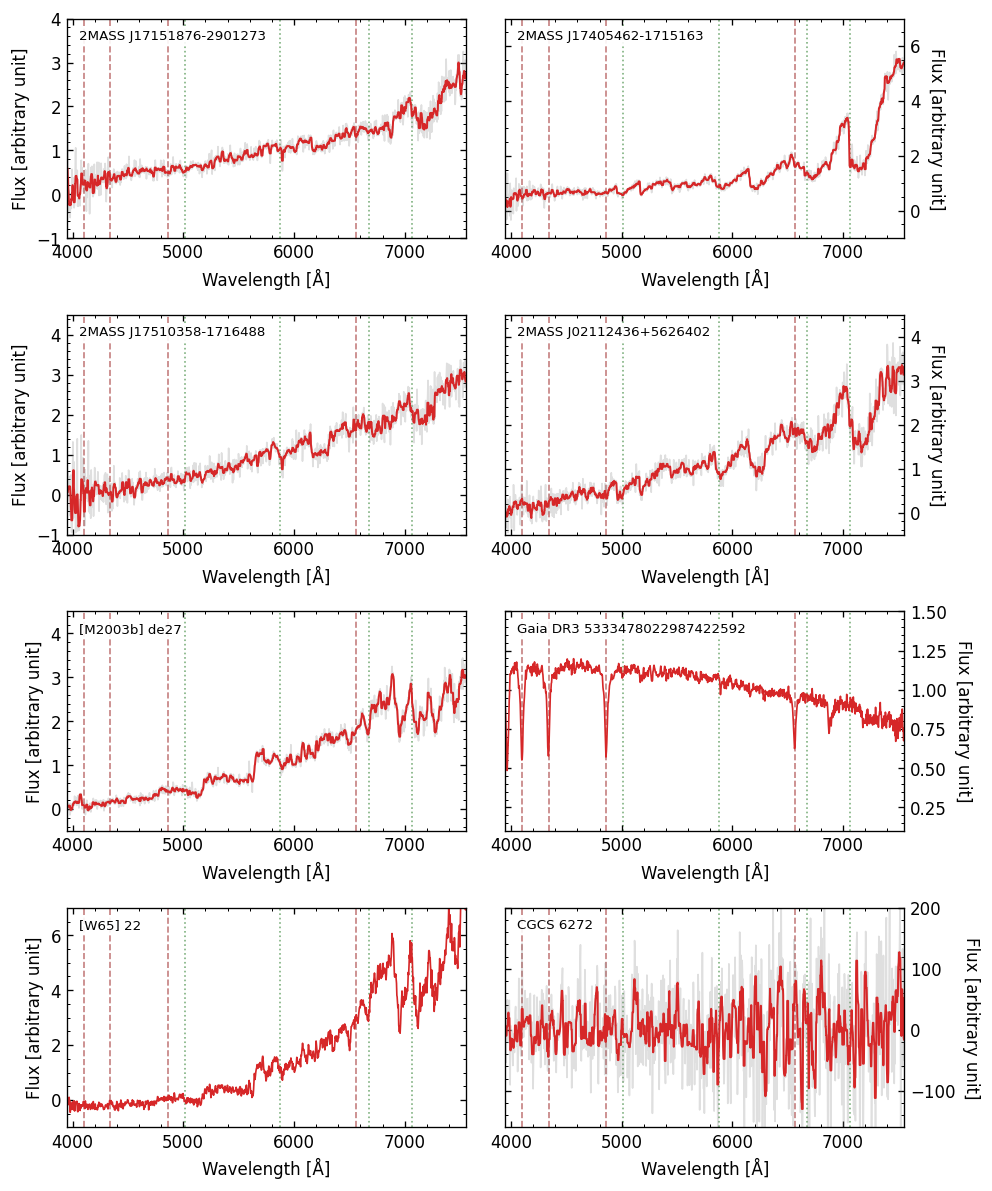}
 \caption{Spectra of eight objects from our control sample, i.e., objects not classified as symbiotics by our pipeline. Expected positions of the Balmer lines are indicated with dashed red lines, while those of \ion{He}{i} lines are marked with dotted green lines.}
 \label{fig:spectra_control}
 \end{figure*}

 \begin{table*}
\centering
\caption{Observing log.}
\label{tab:obslog}
\begin{tabular}{llccc}
\hline
Name & \textit{Gaia} DR3 & Exp. time [s] & JD 24.. & Obs. site \\
\hline

2MASS J14141590-6718454 & 5847837327775225856 & 3 x 1200 & 59844.51 & DSC \\
2MASS J17251717-2948346 & 4059074601049589760 & 6 x 1200 & 59815.59 & DSC \\
V3664 Oph               & 4110870983327739008 & 6 x 1200 & 59814.64 & DSC  \\
PN Sa 3-119             & 4062924712807508736 & 5 x 1200 & 59817.63 & DSC \\
Hen 3-1548              & 4043706834462070784 & 5 x 1200 & 59810.66 & DSC \\
V469 Vul                & 2025674084853852160 & 4 x 1200 & 59786.51 & Ventoux \\
ATO J315.3668+45.9271   & 2163480206474053376 & 5 x 1200 & 59787.50 & OHP \\
2MASS J21180196+5721343 & 2178988199495779456 & 5 x 1200 & 59788.54 & OHP \\
\hline
2MASS J20274687+3031193 & 1861714292419512960 & 5 x 1200 & 61183.57 & Cornillon \\
\hline
2MASS J17151876-2901273 & 4107401714890189312 & 3 x 1200  & 59838.52 & DSC \\
2MASS J17405462-1715163 & 4123903872581171712 & 3 x 1200 & 59838.57 & DSC \\
2MASS J17510358-1716488 & 4144740025867539968 & 3 x 1200 & 59838.62 & DSC \\
2MASS J02112436+5626402  & 457006789709111808  & 5 x 600 & 60181.47 & Ventoux \\
{[}M2003b{]} de27 & 5318385714046320000 & 2 x 600 & 60001.74 & DSC \\
Gaia DR3 5333478022987422592 & 5333478022987422592 & 4 x 600 & 60002.85 & DSC \\
{[}W65{]} 22 & 5339026227414066432 &  5 x 600 & 60022.79 & DSC \\
CGCS 6272 & 5533253788183484672 & 1 x 1000 & 59841.88 & DSC \\
\hline
VPHASDR2 J173155.9-301915.6 & 4058438189969404032 & 4 x 1200 & 61112.86 & DSC \\
Gaia DR3 4064897034844411904 & 4064897034844411904 & 4 x 1200 & 61113.86 & DSC \\
\hline
GSC 00467-02121        & 4268104444583376768 & 5 x 300 & 61154.59 & Calern Observatory \\
2MASS J19245618+2502447 & 2023118300119332608 & 4 x 1200 & 61181.57 & Cornillon \\

\hline
\end{tabular}
\end{table*}


\end{document}